\documentclass{article}

\usepackage{arxiv}

\usepackage[utf8]{inputenc} 
\usepackage[T1]{fontenc}    
\usepackage{hyperref}       
\usepackage{url}            
\usepackage{booktabs}       
\usepackage{amsfonts}       
\usepackage{nicefrac}       
\usepackage{microtype}      
\usepackage{lipsum}
\usepackage{color}
\usepackage{algorithm}
\usepackage{algorithmic}
\usepackage{amsmath,amsfonts,amssymb}
\usepackage{subfigure}
\usepackage{lineno}
\usepackage{graphicx}
\usepackage{bm}
\usepackage{subfigure}

\title{Library-learning-assisted robust principal component analysis for denoising severely corrupted flow fields}

\author{
Pablo Koop$^{[1,2]}$, Isabel Scherl$^{[3]}$, and Kai Fukami$^{[1,*]}$
\\
\\
1. Department of Aerospace Engineering, Graduate School of Engineering, Tohoku University,\\
Sendai, Miyagi 980-8579, Japan\\
2. Faculty of Aerospace Engineering, Delft University of Technology,\\
Kluyverweg 1, 2629 HS Delft, the Netherlands\\
3. Department of Mechanical and Industrial Engineering, University of Massachusetts Amherst, MA 01002, USA\\
~\\
$^{*}$Corresponding author: kfukami1@tohoku.ac.jp
}

\begin{document}
\maketitle

\begin{abstract}

Large-amplitude entrywise corruption can distort flow-field data and contaminate modes extracted from them.
Robust principal component analysis (RPCA) separates low-rank flow content from an entrywise sparse corruption component, but its recovery deteriorates when corruption occupies a large fraction of the measurements.
We introduce library-learning-assisted robust principal component analysis (LLA-RPCA), which restricts the recovered spatial basis to combinations of functions drawn from a fixed candidate library and prescribes the available modal capacity.
Here, the library contains standard trigonometric functions and geometry-adapted graph-Laplacian eigenfunctions, and the resulting decomposition is solved by augmented-Lagrangian alternating minimization.
The method is evaluated using a simulated post-stall NACA0012 wake, video data of an oscillating-cylinder wake, and particle image velocimetry measurements of a flat plate undergoing a transverse extreme gust encounter.
For the first two cases, synthetic large-amplitude entrywise corruption is imposed over fractions ranging from \(0\%\) to \(90\%\).
LLA-RPCA retains coherent wake structures and recovers the dominant proper orthogonal decomposition subspaces at corruption levels where standard RPCA retains residual corruption, attenuates the reconstructed field, or collapses to a one-dimensional reconstruction that no longer represents the time-dependent wake.
For the experimentally measured gust-encounter case, standard RPCA exhibits a trade-off between attenuation of coherent flow content and retention of naturally occurring PIV artifacts as its tuning factor is increased, whereas LLA-RPCA suppresses the artifacts while preserving the coherent velocity and derived-vorticity structures more consistently.
These results indicate that a library-constrained reconstruction with prescribed modal capacity can improve denoising and modal recovery when the prescribed representation adequately captures the relevant spatial content.

\end{abstract}

\section{Introduction}
\label{sec:intro}

Robust principal component analysis (RPCA) provides a global matrix-based approach for denoising data by decomposing the observed snapshot matrix into a low-rank component used to represent the coherent content and an entrywise sparse corruption component \cite{Candes2011-RPCA,huang2012singing,bouwmans2014robust}.
Scherl et al.~\cite{scherl2020robust} demonstrated that RPCA can be applied as a preprocessing filter to numerical and experimental flow-field data containing artificial outliers and measurement corruption.
Compared with POD and DMD applied directly to the corrupted data, RPCA filtering can improve the recovery of dominant modal structures and temporal dynamics~\cite{scherl2020robust}.
RPCA therefore provides an appropriate baseline for reconstructing flow when the dominant flow content is approximately low-rank and large-amplitude entrywise corruption affects only a limited fraction of the snapshot matrix.

However, reliable separation of coherent flow content and corruption with RPCA becomes more difficult as the corruption fraction increases.
Theoretical recovery guarantees require the spatial and temporal directions of the low-rank component not to be concentrated on only a small set of coordinates, while the corruption must affect a sufficiently limited set of entries and must not itself form a low-rank pattern \cite{Candes2011-RPCA}.
As the corruption fraction increases, the corruption becomes less compatible with the assumed sparse support, while fewer uncorrupted entries remain to constrain the separation of the two components \cite{Candes2011-RPCA,bouwmans2018applications}.
Part of the corruption may remain in the low-rank component, producing a contaminated reconstruction.
In contrast, physical flow content may be assigned to the sparse component, producing attenuation or loss of coherent structures.
The intended decomposition should therefore avoid both contamination of the recovered low-rank component and removal of physical flow content.
These allocation ambiguities are not restricted to globally high corruption fractions, since strongly corrupted individual snapshots or spatially concentrated artifacts can likewise impose competing requirements on the separation of physical and corrupted content.

Physical-content transfer can reduce the nuclear norm by attenuating singular contributions without necessarily changing the rank.
If this transfer is sufficiently strong to eliminate singular directions, however, the rank of the recovered low-rank component also decreases.
Because standard RPCA does not prescribe this rank directly, such rank loss can leave too few spatial directions to represent all relevant coherent structures.
Changing the RPCA hyperparameter modifies the balance between the low-rank and sparse components and can reduce physical-content transfer, but can also allow more corrupted content to remain in the reconstruction.
It therefore does not provide independent control over the modal capacity of the recovered flow while maintaining effective corruption removal.

To address these limitations, we propose library-learning-assisted robust principal component analysis (LLA-RPCA).
The method introduces two main modifications to the recovered low-rank representation.
First, its spatial basis is learned from a prescribed candidate library, which restricts the spatial patterns that can be represented in the reconstructed flow.
When the candidate library represents coherent flow structures more effectively than the irregular deviations introduced by corrupted entries, this restriction reduces the ability of the recovered component to represent corruption.
Second, the recovered component is represented explicitly using a prescribed number of learned spatial modes rather than through nuclear-norm minimization.
This removes the direct nuclear-norm incentive to attenuate physical content and provides control over the modal capacity available to represent the coherent flow.

We evaluate LLA-RPCA using three datasets that cover different aspects of the reconstruction problem.
For the first case, we consider a computational data set of the two-dimensional laminar post-stall wake behind a NACA0012 airfoil at \(\alpha=40^\circ\) at a chord-based Reynolds number of ${\rm Re}=100$ \cite{cliff1,cliff2,fukami2023grasping}.
This case provides a representation of corrupted PIV data for which an exact uncorrupted reference is available.
We impose controlled large-amplitude entrywise corruption using two spatial distributions.
The first is random, while the second is concentrated in regions of high shear, where PIV measurements are more prone to error.
The original fields are retained as an uncorrupted reference for quantitative evaluation.
The second case considers an experimental image-intensity sequence of vortex shedding behind a cylinder oscillating in the streamwise direction, obtained from the APS Gallery of Fluid Motion \cite{boersma2021vortexarms}.
Although the cylinder motion is periodic, the observed wake does not repeat exactly between successive oscillation cycles, with variations in the streakline and vortex patterns producing quasi-periodic flow evolution.
We use this case to examine whether the reconstruction performance of LLA-RPCA extends to different flow physics with less regular temporal behavior.
We again impose controlled large-amplitude entrywise corruption while retaining the original image sequence as an uncorrupted reference.

For the third case, we evaluate the proposed method on experimentally acquired PIV velocity fields containing naturally occurring artifacts rather than artificially imposed corruption.
The data are obtained from time-resolved measurements of a flat plate at \(\alpha=0^\circ\) towed through a transverse jet at ${\rm Re}=20{,}000$ \cite{biler2021experimental,towne2023database}.
RPCA and LLA-RPCA are applied directly to the measured streamwise and transverse velocity components, where the PIV artifacts are present, before spanwise vorticity is calculated from the reconstructed fields.
This ordering suppresses the measurement artifacts at the velocity-field level before spatial differentiation can amplify them in the derived vorticity.
The case also enables the sensitivity of standard RPCA to its tuning factor to be examined using snapshots with substantially different levels of visible PIV artifacts.
Together, the three cases evaluate whether LLA-RPCA reduces corrupted content retained in the recovered flow fields while preserving physical flow content, coherent spatial structures, and the modal capacity required to represent the underlying dynamics.
The present paper is organized as follows: the method of the proposed LLA-RPCA is described in section~\ref{sec:method}.
We discuss the reconstruction results with three flow examples in section~\ref{sec:results}.
Conclusions are finally remarked in section~\ref{sec:conc}.

\section{LLA-RPCA methodology}
\label{sec:method}

Standard RPCA decomposes the observed snapshot matrix \(\mathbf{X}\in\mathbb{R}^{m\times n}\) into a
low-rank component \(\mathbf{L}\) and an entrywise sparse component \(\mathbf{S}\) by
solving~\cite{Candes2011-RPCA,scherl2020robust}

\begin{equation}
    \min_{\mathbf{L},\mathbf{S}}
    \|\mathbf{L}\|_{*}
    +
    \frac{\lambda_{\mathrm{RPCA}}}{\sqrt{\max(m,n)}}\|\mathbf{S}\|_{1}
    \quad \text{s.t.} \quad
    \mathbf{X} = \mathbf{L} + \mathbf{S},
    \label{eq:standard_rpca}
\end{equation}

where \(\|\mathbf{L}\|_*\) is the nuclear norm, \(\|\mathbf{S}\|_1\) is the entrywise \(\ell_1\)
norm, and \(\lambda_{\mathrm{RPCA}}\) is a dimensionless tuning factor controlling the balance
between the two penalties.
Following Scherl et al.~\cite{scherl2020robust}, \(\lambda_{\mathrm{RPCA}}=1\) is used for the two controlled validation datasets.
For the experimental flat-plate case, \(\lambda_{\mathrm{RPCA}}=1\) and \(1.4\) are considered to
examine the balance between preservation of coherent flow content and suppression of naturally
occurring PIV artifacts.
Because the number of retained spatial locations exceeds the number of snapshots in all validation datasets, the
coefficient multiplying \(\|\mathbf{S}\|_1\) reduces to
\(\lambda_{\mathrm{RPCA}}/\sqrt{m}\).

The nuclear norm promotes a low-rank reconstruction but does not prescribe which low-rank spatial subspace should represent the physical flow, while the entrywise \(\ell_1\) norm promotes sparsity without identifying which observed content should belong to the corruption component.
The equality constraint fixes only the sum of the two components, so the balance between these penalties can favor two opposing misallocations.
First, corrupted content can remain in \(\mathbf{L}\), producing contamination of the recovered flow field.
Second, physical flow content can be transferred from \(\mathbf{L}\) to \(\mathbf{S}\) when the resulting reduction in nuclear norm outweighs the associated increase in the weighted \(\ell_1\) penalty, producing attenuation.
This attenuation does not require a reduction in rank because reducing nonzero singular contributions is sufficient to decrease the nuclear norm.
If physical-content transfer is sufficiently strong to eliminate singular directions, however, the rank also decreases.
Because standard RPCA does not prescribe the rank of \(\mathbf{L}\) directly, this can leave insufficient modal capacity to represent the coherent flow.
Increasing \(\lambda_{\mathrm{RPCA}}\) makes assignment of content to \(\mathbf{S}\) more expensive and can therefore reduce physical-content transfer, but can simultaneously favor retention of corrupted content in \(\mathbf{L}\).
The tuning factor consequently controls a trade-off between the two failure modes rather than providing independent control over corruption removal, attenuation, and modal capacity.
LLA-RPCA addresses the absence of spatial structural information by representing the recovered
flow component using \(r\) learned spatial modes formed from a prescribed candidate library.
The recovered low-rank matrix is parameterized as

\begin{equation}
    \label{eq:low_rank_parameterization}
    \mathbf{L} = \boldsymbol{\Psi}\mathbf{B}\mathbf{A}.
\end{equation}

Here, \(\boldsymbol{\Psi}\in\mathbb{R}^{m\times k}\) is the fixed candidate library, where \(k\) is
the number of retained candidate-library directions. The matrix
\(\mathbf{B}\in\mathbb{R}^{k\times r}\) weights and combines these directions to form the learned
spatial basis

\begin{equation}
    \boldsymbol{\Phi} = \boldsymbol{\Psi}\mathbf{B} \in \mathbb{R}^{m\times r}.
\end{equation}

The columns of \(\boldsymbol{\Phi}\) are the \(r\) learned spatial modes used to reconstruct the coherent flow from the corrupted observations.
The matrix \(\mathbf{A}\in\mathbb{R}^{r\times n}\) contains their snapshot-dependent modal coefficients.
The recovered low-rank matrix can therefore also be written as $\mathbf{L} = \boldsymbol{\Phi}\mathbf{A}$.
The candidate library determines which shared spatial directions may be represented in \(\mathbf{L}\), while \(r\) specifies the modal capacity used to describe coherent variation across the snapshot sequence.
When the library represents the coherent flow more effectively than the irregular deviations introduced by corruption, the candidate-library restriction reduces the ability of the recovered component to reproduce corrupted patterns.
The explicit factorization simultaneously replaces the nuclear-norm-based determination of the low-rank representation used in standard RPCA~\cite{scherl2020robust,Candes2011-RPCA}.
Consequently, LLA-RPCA does not contain a nuclear-norm penalty that can favor attenuation of physical content solely to reduce the nuclear norm, while the prescribed value of \(r\) directly controls the modal capacity available to represent the coherent flow.
The columns of \(\mathbf{A}\) are not constrained to vary smoothly from one snapshot to the next.
The formulation therefore restricts the shared spatial representation without imposing temporal regularity or a dynamical model.
This permits transient and nonperiodic evolution to be represented, but it also allows an individual coefficient vector to become poorly constrained when too few informative entries remain in a snapshot.

\begin{figure*}[t]
    \centering
    \includegraphics[width=\textwidth]{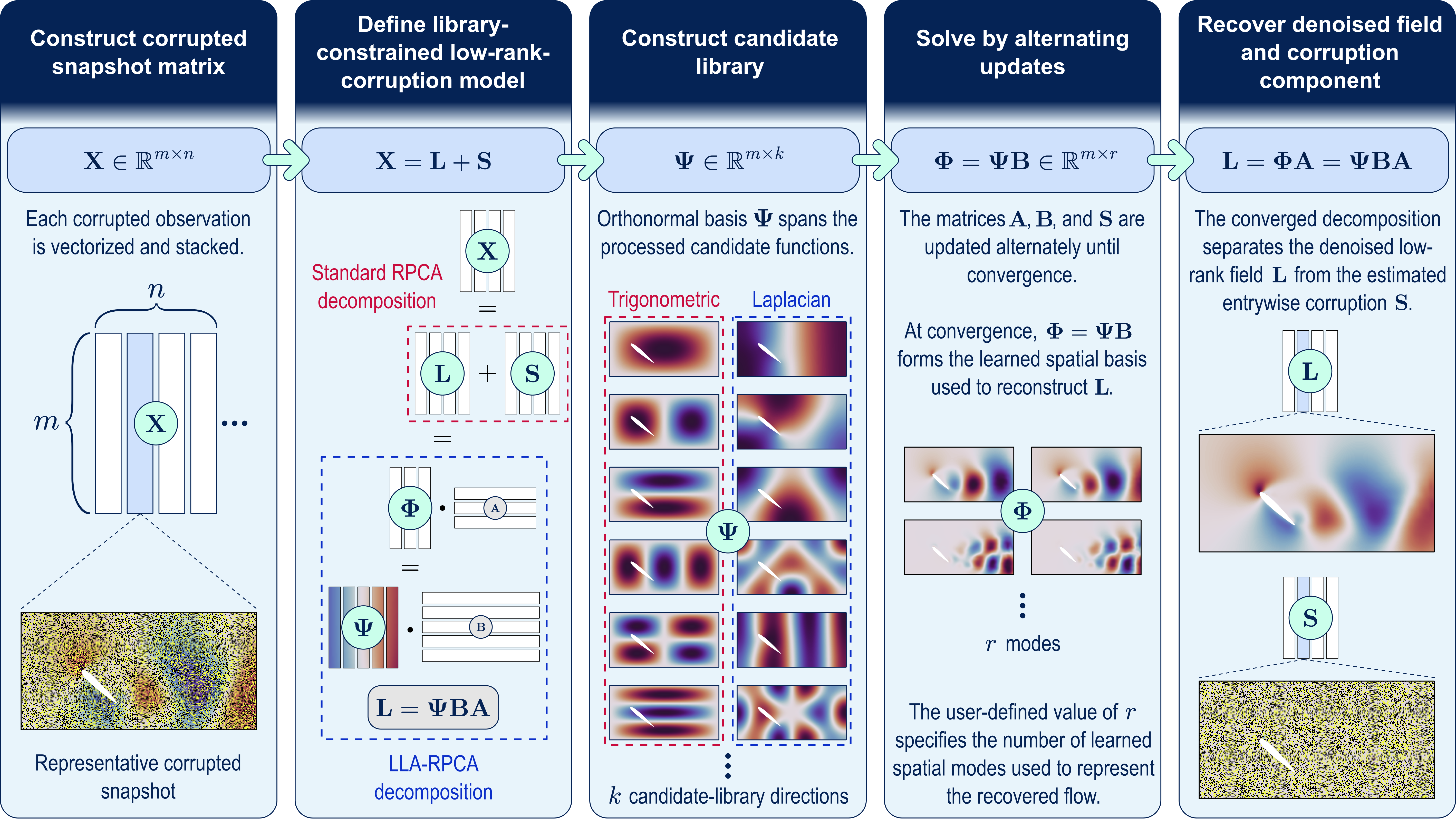}
    \caption{
        Overview of the LLA-RPCA workflow. The observed matrix \(\mathbf{X}\) is decomposed
        as \(\mathbf{X}=\boldsymbol{\Psi}\mathbf{B}\mathbf{A}+\mathbf{S}\), where
        \(\boldsymbol{\Psi}\) is the prescribed candidate library, \(\mathbf{B}\) forms the
        learned spatial basis \(\boldsymbol{\Phi}=\boldsymbol{\Psi}\mathbf{B}\),
        \(\mathbf{A}\) contains its snapshot-dependent modal coefficients, and
        \(\mathbf{S}\) is the entrywise corruption component. The denoised field is
        recovered as \(\mathbf{L}=\boldsymbol{\Psi}\mathbf{B}\mathbf{A}\). The user-defined
        value \(r\) specifies the modal capacity used to represent the recovered flow.
    }
    \label{fig:method_overview}
    \vspace{-2mm}
\end{figure*}

Figure~\ref{fig:method_overview} summarizes the LLA-RPCA workflow.
The candidate library
\(\boldsymbol{\Psi}\) is constructed first, after which the constrained decomposition is solved
through alternating updates of \(\mathbf{A}\), \(\mathbf{B}\), and \(\mathbf{S}\). These updates
determine the learned basis \(\boldsymbol{\Phi}=\boldsymbol{\Psi}\mathbf{B}\) and the corruption
component \(\mathbf{S}\). After convergence, \(\mathbf{L}=\boldsymbol{\Phi}\mathbf{A}\) is reshaped
into denoised flow-field snapshots.
\vspace{-5mm}
\subsection{Library construction}
\label{subsec:library_construction}

The candidate library $\boldsymbol{\Psi}$ determines the admissible spatial subspace from which the recovered low-rank field can be constructed.
The formulation permits any prescribed candidate functions suited to the expected coherent structures, including analytical, geometry-adapted, and data-driven spatial patterns.
The present library combines trigonometric functions with graph-Laplacian eigenfunctions.
The number of candidates retained from each family and the prescribed modal capacity are case-dependent modeling choices and are reported for each validation case.
The trigonometric candidates provide spatial patterns with progressively increasing spatial variation.
For a two-dimensional domain with coordinates $(x,y)$ and characteristic dimensions $L_x$ and $L_y$, the default candidates are separable sine products,

\begin{equation}
    \psi_{pq}^{\mathrm{trig}}(x,y) = \sin\left(\frac{p\pi x}{L_x}\right) \sin\left(\frac{q\pi
        y}{L_y}\right), \qquad p=1,\ldots,p_{\max}, \qquad q=1,\ldots,q_{\max}.
    \label{eq:trigonometric_library}
\end{equation}

Low-order combinations describe large-scale spatial variations, whereas increasing $p$ and $q$
introduces progressively finer structures. These functions are defined on the rectangular
computational grid and therefore do not directly account for immersed solid geometries. To provide
geometry-adapted candidates, we additionally use eigenfunctions of the discrete graph Laplacian
constructed on the valid fluid cells,

\begin{equation}
    \mathbf{L}_{\Delta} \boldsymbol{\psi}_{\ell}^{\mathrm{Lap}} = \xi_{\ell}
        \boldsymbol{\psi}_{\ell}^{\mathrm{Lap}},
    \label{eq:laplacian_library}
\end{equation}

where $\mathbf{L}_{\Delta}$ is the graph Laplacian on the valid-fluid grid. Connections crossing into masked regions are excluded, so the resulting eigenfunctions adapt to the immersed geometry.
Eigenfunctions associated with small eigenvalues vary slowly between neighboring fluid cells, whereas larger eigenvalues correspond to progressively finer spatial variations.
Before assembling the candidate library, the candidates are restricted to the valid fluid domain and processed with respect to their spatial means.
The nonconstant candidates are spatially mean-centered.
A separate spatially constant candidate is retained for the signed NACA0012 data and the experimental flat-plate data to represent snapshot-dependent variations in the spatial mean.
For validation of the NACA0012 case, approximately equal numbers of selected entries are replaced by \(+10\sigma\) and \(-10\sigma\), consistent with the signed nature of the flow variable.
The imposed replacement values therefore contain no global sign bias, although the resulting corruption relative to the clean field is not necessarily zero-mean at each spatial location because the original values are replaced rather than perturbed additively.
In contrast, selected entries of the non-negative cylinder intensity data are replaced by positive values only.
For the fixed-amplitude replacement model used here, high corruption fractions can therefore make the observed matrix strongly dominated by the repeated positive replacement value and an associated spatially constant low-rank pattern.
If a constant candidate were included in the cylinder library, LLA-RPCA could represent this pattern directly and could therefore absorb the dominant positive corruption into the recovered low-rank field.
To exclude this failure route by modeling choice, no separate constant candidate is retained for the cylinder case, while all retained nonconstant candidates are spatially mean-centered.
A nonzero spatially constant field is therefore outside the admissible LLA-RPCA spatial subspace.
An important consideration in constructing the library is the initialization of the learned spatial basis $\boldsymbol{\Phi}^{(0)}$. 
In the decomposition $\mathbf{L}=\boldsymbol{\Psi}\mathbf{B}\mathbf{A}$, the matrix $\mathbf{B}$ combines the library directions in $\boldsymbol{\Psi}$ into the learned spatial basis $\boldsymbol{\Phi}$.
Because $\mathbf{B}$ and the snapshot-dependent modal coefficient matrix $\mathbf{A}$ enter through the bilinear product $\mathbf{B}\mathbf{A}$, the optimization is jointly nonconvex. 
Different initial learned subspaces may therefore lead to different converged decompositions, and recovery of the globally optimal decomposition is not guaranteed.
We choose $\boldsymbol{\Phi}^{(0)}$ to limit the amount of corruption represented in the first low-rank estimate, because corruption absorbed at this stage is less likely to be isolated in the corruption component $\mathbf{S}$ during the subsequent updates, as discussed in section~\ref{subsubsec:update_S}.
We therefore form $\boldsymbol{\Phi}^{(0)}$ from the smoothest independent directions in the candidate library $\boldsymbol{\Psi}$. 
Linear combinations of these smooth directions remain spatially broad and have limited ability to reproduce isolated corrupted entries.
By contrast, combinations of rougher directions can reinforce locally while canceling elsewhere, making localized corruption easier to represent in the initial low-rank estimate.
A common smoothness measure is therefore required to rank the trigonometric, graph-Laplacian, and any other included candidate families.
For a processed candidate $\widetilde{\boldsymbol{\psi}}_j$, its spatial roughness is defined by the graph-Dirichlet Rayleigh quotient,

\begin{equation}
    \mathcal{R}_j = \frac{ \widetilde{\boldsymbol{\psi}}_j^{T} \mathbf{L}_{\Delta}
        \widetilde{\boldsymbol{\psi}}_j }{ \widetilde{\boldsymbol{\psi}}_j^{T}
        \widetilde{\boldsymbol{\psi}}_j }.
    \label{eq:library_roughness}
\end{equation}

When each undirected graph edge is counted once, this quantity can equivalently be written as

\begin{equation}
    \mathcal{R}_j = \frac{ \displaystyle \sum_{(a,b)\in\mathcal{E}} \left[ \widetilde{\psi}_j(a) -
        \widetilde{\psi}_j(b) \right]^2 }{ \displaystyle \sum_{a\in\mathcal{V}}
        \widetilde{\psi}_j(a)^2 },
    \label{eq:library_roughness_edges}
\end{equation}

where $\mathcal{V}$ denotes the valid fluid cells and $\mathcal{E}$ denotes neighboring valid-cell
pairs. The numerator measures variation between adjacent cells, while the denominator removes
dependence on the amplitude of the candidate. Small values of $\mathcal{R}_j$ correspond to smooth
spatial structures, whereas large values indicate rapidly varying structures. For a graph-Laplacian
eigenfunction, equation~\eqref{eq:laplacian_library} gives

\begin{equation}
    \mathcal{R}_j = \xi_j.
    \label{eq:laplacian_roughness_eigenvalue}
\end{equation}

The same measure provides a geometry-aware estimate of the spatial variation of the masked
trigonometric candidates, enabling all candidate families to be ranked using one common criterion.
The processed candidates are sorted in increasing roughness before the complete library is
orthonormalized,

\begin{equation}
    \mathcal{R}_{\pi_1} \leq \mathcal{R}_{\pi_2} \leq \cdots \leq \mathcal{R}_{\pi_k},
    \label{eq:roughness_order}
\end{equation}

and assembled into the ordered candidate matrix

\begin{equation}
    \widetilde{\boldsymbol{\Psi}} = \begin{bmatrix} \widetilde{\boldsymbol{\psi}}_{\pi_1} &
        \widetilde{\boldsymbol{\psi}}_{\pi_2} & \cdots & \widetilde{\boldsymbol{\psi}}_{\pi_k}
        \end{bmatrix}.
    \label{eq:roughness_ordered_library}
\end{equation}

The reordering does not change the full subspace spanned by the candidate set. It determines the
sequence in which independent spatial directions are introduced during the subsequent
orthonormalization. This sequence is relevant because the leading independent directions are used to
initialize the learned basis. Before initialization, the ordered candidate matrix is
orthonormalized. The processed candidates are not mutually orthogonal over the valid fluid domain
because masking destroys the orthogonality of the trigonometric candidates, while candidates from
different families may contain overlapping directions. The main reason for orthonormalizing the
library is that the relation $\boldsymbol{\Psi}^{T}\boldsymbol{\Psi}=\mathbf{I}_k$ simplifies the
update of the candidate-combination coefficient matrix $\mathbf{B}$, as shown in
Appendix~\ref{app:update_B}. Without this relation, the update remains coupled through
$\boldsymbol{\Psi}^{T}\boldsymbol{\Psi}$, requiring the solution of a linear system whose dimension
scales with the library size \(k\). For an orthonormal library, this coupling disappears, and the
matrix inversion is reduced to an $r\times r$ system that is independent of the library size.
Orthonormalization replaces the correlated candidate coordinates by orthonormal coordinates spanning
the same retained spatial subspace, thereby improving the conditioning of the optimization. We
therefore replace the ordered candidate matrix by an orthonormal basis for the same spatial subspace
using an unpivoted QR factorization.

\begin{equation}
    \widetilde{\boldsymbol{\Psi}} = \mathbf{Q}_{\Psi}\mathbf{R}_{\Psi}.
    \label{eq:library_qr_factorization}
\end{equation}

The library used during the optimization is defined as

\begin{equation}
    \boldsymbol{\Psi} = \mathbf{Q}_{\Psi}, \qquad \boldsymbol{\Psi}^{T}\boldsymbol{\Psi} =
        \mathbf{I}_k.
    \label{eq:orthonormal_library_definition}
\end{equation}

Because the QR factorization is performed without column pivoting, the first $r$ columns of
$\boldsymbol{\Psi}$ span the same subspace as the first $r$ retained roughness-ordered candidate
directions. The leading columns of $\boldsymbol{\Psi}$ therefore provide an orthonormal
representation of the smooth candidate subspace used to initialize the learned basis. We initialize
the spatial coefficient matrix as

\begin{equation}
    \mathbf{B}^{(0)} = \begin{bmatrix} \mathbf{I}_{r} \\ \mathbf{0} \end{bmatrix} \in
        \mathbb{R}^{k\times r},
    \label{eq:B_initialization}
\end{equation}

which gives the initial learned basis

\begin{equation}
    \boldsymbol{\Phi}^{(0)} = \boldsymbol{\Psi}\mathbf{B}^{(0)} = \boldsymbol{\Psi}_{:,1:r}.
    \label{eq:Phi_initialization}
\end{equation}

The initial active subspace is thus an orthonormal representation of the subspace generated by the
smoothest retained independent candidate directions. The remaining optimization variables are
initialized as

\begin{equation}
    \mathbf{A}^{(0)} = \mathbf{0}, \qquad \mathbf{S}^{(0)} = \mathbf{0}, \qquad \mathbf{Y}^{(0)} =
        \mathbf{0}.
    \label{eq:remaining_initialization}
\end{equation}

The initial learned basis is selected independently of the observed snapshot matrix $\mathbf{X}$,
using only the prescribed library and its roughness ordering. It determines the starting subspace of
the optimization but does not constrain the final reconstruction to the smoothest library
directions. During the optimization, the entries of $\mathbf{B}$ are updated freely, allowing the
learned basis to combine all available library directions, including finer-scale candidates required
to represent coherent flow structures.
\vspace{-2mm}
\subsection{Optimization problem and augmented Lagrangian}
\label{subsec:optimization_problem}

We formulate the library-constrained decomposition as a constrained optimization problem. The
complete constrained optimization problem is

\begin{equation}
    \label{eq:constrained_problem}
    \min_{\mathbf{A},\mathbf{B},\mathbf{S}} \|\mathbf{S}\|_1 \quad \text{s.t.} \quad \mathbf{X} =
        \boldsymbol{\Psi}\mathbf{B}\mathbf{A} + \mathbf{S}.
\end{equation}

Unlike standard RPCA, equation~\eqref{eq:constrained_problem} contains no nuclear-norm penalty on
$\mathbf{L}$ because the admissible low-rank representation is imposed directly through the prescribed modal capacity $r$ and the spatial subspace of the candidate library $\boldsymbol{\Psi}$.
We solve this constrained problem using an augmented Lagrangian approach. The decomposition
requirement forms the equality constraint
$\mathbf{X}=\boldsymbol{\Psi}\mathbf{B}\mathbf{A}+\mathbf{S}$. Following the standard augmented
Lagrangian construction for equality-constrained optimization problems~\cite{nocedal2006numerical},
we write

\begin{align}
    \mathcal{L}_{\nu}(\mathbf{A},\mathbf{B},\mathbf{S},\mathbf{Y}) ={}& \|\mathbf{S}\|_1 +
        \left\langle \mathbf{Y}, \mathbf{X} - \boldsymbol{\Psi}\mathbf{B}\mathbf{A} - \mathbf{S}
        \right\rangle \nonumber\\
    & + \frac{\nu}{2} \left\| \mathbf{X} - \boldsymbol{\Psi}\mathbf{B}\mathbf{A} - \mathbf{S}
        \right\|_F^2,
    \label{eq:augmented_lagrangian_initial}
\end{align}

where $\mathbf{Y}$ is the Lagrange multiplier matrix associated with the equality constraint and
$\nu>0$ is the augmented-Lagrangian penalty parameter. At iteration $h$, the penalty parameter is
denoted by $\nu^{(h)}$ and is increased as

\begin{equation}
    \nu^{(h+1)} = \min \left( \gamma_{\nu}\nu^{(h)}, \nu_{\max} \right), \qquad \gamma_{\nu}>1,
    \label{eq:nu_continuation}
\end{equation}

where $\nu^{(0)}$ is the initial penalty value, $\gamma_{\nu}$ is the multiplicative growth factor,
and $\nu_{\max}$ is the maximum allowed penalty value. For the simulated post-stall NACA0012 wake
dataset with $\eta=0.4$, the reconstruction error changed only marginally when $\nu^{(0)}$ was
varied over the logarithmically spaced values from $10^{-3}$ to $10^{2}$. Based on this sensitivity
assessment, we use $\nu^{(0)}=0.01$ for all three validation datasets and set $\gamma_{\nu}=1.05$
and $\nu_{\max}=10^{6}$ throughout. Increasing $\nu^{(h)}$ progressively strengthens the penalty on
violations of the decomposition constraint. The inner-product term represents the Lagrange
multiplier contribution enforcing the structural decomposition, while the quadratic term penalizes
violations of the equality constraint. This formulation adapts standard augmented Lagrangian
methods~\cite{lin2010augmented} to accommodate the explicit library-parameterized form of the
low-rank component. Completing the square combines the multiplier and penalty terms. After
discarding constants that are independent of the primal variables $\mathbf{A}$, $\mathbf{B}$, and
$\mathbf{S}$, the reduced augmented Lagrangian used for the optimization updates is

\begin{equation}
    \mathcal{L}_{\nu}^{\mathrm{red}} (\mathbf{A},\mathbf{B},\mathbf{S},\mathbf{Y}) =
        \|\mathbf{S}\|_1 + \frac{\nu}{2} \left\| \mathbf{X} - \boldsymbol{\Psi}\mathbf{B}\mathbf{A}
        - \mathbf{S} + \frac{1}{\nu}\mathbf{Y} \right\|_F^2.
    \label{eq:augmented_lagrangian_reduced}
\end{equation}

At iteration $h$, this expression is evaluated using $\nu=\nu^{(h)}$.
Appendix~\ref{app:reduced_augmented_lagrangian} provides the derivation of this reduced form.
\vspace{-2mm}
\subsection{Alternating minimization sequence}
\label{subsec:alternating_minimization}

Because the product \(\mathbf{B}\mathbf{A}\) is bilinear, the optimization is jointly nonconvex in
\(\mathbf{A}\) and \(\mathbf{B}\). We therefore update \(\mathbf{A}\), \(\mathbf{B}\),
\(\mathbf{S}\), and \(\mathbf{Y}\) sequentially, with a QR reparameterization after the
\(\mathbf{B}\)-update. Each block subproblem is solved exactly for the current values of the
remaining variables. The converged decomposition is not guaranteed to be globally optimal and may
depend on the initialization, candidate library, prescribed modal capacity, and continuation
schedule. This dependence motivates the smooth-subspace initialization described in
section~\ref{subsec:library_construction}. The full derivations of the $\mathbf{A}$-, $\mathbf{B}$-,
and $\mathbf{S}$-updates are provided in Appendix~\ref{app:alternating_updates}.
\vspace{-2mm}
\subsubsection{Update for snapshot-dependent modal coefficients $\mathbf{A}$}
\label{subsubsec:update_A}

We first update the snapshot-dependent modal coefficient matrix $\mathbf{A}$ while holding
$\mathbf{B}$, $\mathbf{S}$, and $\mathbf{Y}$ fixed. Let
$\boldsymbol{\Phi}^{(h)}=\boldsymbol{\Psi}\mathbf{B}^{(h)}$ denote the learned basis at iteration
$h$. Because the learned basis has orthonormal columns after initialization and the QR
reparameterization, $(\boldsymbol{\Phi}^{(h)})^T\boldsymbol{\Phi}^{(h)}=\mathbf{I}_r$ in the present
implementation. The closed-form solution to this least-squares subproblem is obtained by setting the
gradient of the augmented Lagrangian with respect to $\mathbf{A}$ to zero,
$\nabla_{\mathbf{A}}\mathcal{L}_{\nu^{(h)}}=\mathbf{0}$:

\begin{equation}
    \label{eq:update_A_main}
    \mathbf{A}^{(h+1)} = (\boldsymbol{\Phi}^{(h)})^T \left( \mathbf{X} - \mathbf{S}^{(h)} +
        \frac{1}{\nu^{(h)}}\mathbf{Y}^{(h)} \right).
\end{equation}
\vspace{-6mm}
\subsubsection{Update for mode combination coefficients $\mathbf{B}$}
\label{subsubsec:update_B}

Next, we update the mode-combination matrix $\mathbf{B}$ while holding $\mathbf{A}$, $\mathbf{S}$,
and $\mathbf{Y}$ fixed. Setting the gradient of the augmented Lagrangian with respect to
\(\mathbf{B}\) to zero and using \(\boldsymbol{\Psi}^{T}\boldsymbol{\Psi}=\mathbf{I}_k\) gives

\begin{equation}
    \mathbf{B}^{(h+1)} = \boldsymbol{\Psi}^{T} \left( \mathbf{X} - \mathbf{S}^{(h)} +
        \frac{1}{\nu^{(h)}}\mathbf{Y}^{(h)} \right) (\mathbf{A}^{(h+1)})^{T} \left[
        \mathbf{A}^{(h+1)} (\mathbf{A}^{(h+1)})^{T} \right]^{-1}.
    \label{eq:update_B_main}
\end{equation}

The orthonormality of $\boldsymbol{\Psi}$ removes the coupling through
$\boldsymbol{\Psi}^{T}\boldsymbol{\Psi}$ from the normal equations. Without this simplification, the
update requires solving a coupled system whose dimension scales with the library size \(k\), as
mentioned above. For $\boldsymbol{\Psi}^{T}\boldsymbol{\Psi}=\mathbf{I}_k$, the update reduces to
the inversion of the $r\times r$ matrix in equation~\eqref{eq:update_B_main}, which is independent
of the library size \(k\).
\vspace{-4mm}
\subsubsection{Learned basis orthogonalization}
\label{subsubsec:orthogonalization}

The learned basis $\boldsymbol{\Phi}=\boldsymbol{\Psi}\mathbf{B}$ does not necessarily remain
orthogonal during the alternating updates. Although the candidate library $\boldsymbol{\Psi}$ is
orthonormal after preprocessing, arbitrary combinations of its columns through $\mathbf{B}$ can
produce nearly parallel learned modes. This can make the subsequent temporal-coefficient update
poorly conditioned. We therefore orthogonalize the learned basis during each iteration. We first
apply a reduced QR factorization to the current learned basis,

\begin{equation}
    \boldsymbol{\Psi}\mathbf{B}^{(h+1)} = \mathbf{Q}_{\Phi}^{(h+1)} \mathbf{R}_{\Phi}^{(h+1)},
\end{equation}

where the columns of $\mathbf{Q}_{\Phi}^{(h+1)}$ are orthonormal. Provided that
$\boldsymbol{\Psi}\mathbf{B}^{(h+1)}$ has full column rank, $\mathbf{R}_{\Phi}^{(h+1)}$ is
nonsingular. We then define the reparameterized coefficient matrices as

\begin{equation}
    \widehat{\mathbf{B}}^{(h+1)} = \mathbf{B}^{(h+1)} \left( \mathbf{R}_{\Phi}^{(h+1)} \right)^{-1},
        \qquad \widehat{\mathbf{A}}^{(h+1)} = \mathbf{R}_{\Phi}^{(h+1)} \mathbf{A}^{(h+1)}.
\end{equation}

The learned basis associated with $\widehat{\mathbf{B}}^{(h+1)}$ is

\begin{align}
    \boldsymbol{\Psi}\widehat{\mathbf{B}}^{(h+1)} &= \boldsymbol{\Psi}\mathbf{B}^{(h+1)} \left(
        \mathbf{R}_{\Phi}^{(h+1)} \right)^{-1} \nonumber\\
    &= \mathbf{Q}_{\Phi}^{(h+1)} \mathbf{R}_{\Phi}^{(h+1)} \left( \mathbf{R}_{\Phi}^{(h+1)}
        \right)^{-1} \nonumber\\
    &= \mathbf{Q}_{\Phi}^{(h+1)}.
\end{align}

The reparameterized learned basis therefore has orthonormal columns,

\begin{equation}
    \left( \boldsymbol{\Psi}\widehat{\mathbf{B}}^{(h+1)} \right)^T \left(
        \boldsymbol{\Psi}\widehat{\mathbf{B}}^{(h+1)} \right) = \mathbf{I}_r.
\end{equation}

At the same time, the reconstructed low-rank matrix remains unchanged because

\begin{align}
    \boldsymbol{\Psi} \widehat{\mathbf{B}}^{(h+1)} \widehat{\mathbf{A}}^{(h+1)} &= \boldsymbol{\Psi}
        \left[ \mathbf{B}^{(h+1)} \left( \mathbf{R}_{\Phi}^{(h+1)} \right)^{-1} \right] \left[
        \mathbf{R}_{\Phi}^{(h+1)} \mathbf{A}^{(h+1)} \right] \nonumber\\
    &= \boldsymbol{\Psi} \mathbf{B}^{(h+1)} \mathbf{A}^{(h+1)}.
\end{align}

The QR step therefore changes only the internal factorization of the current low-rank matrix. It
replaces the learned spatial basis by an orthonormal basis for the same subspace and transfers the
corresponding scaling and rotation into the modal coefficients. After this reparameterization, we
set

\begin{equation}
    \mathbf{B}^{(h+1)} \leftarrow \widehat{\mathbf{B}}^{(h+1)}, \qquad \mathbf{A}^{(h+1)} \leftarrow
        \widehat{\mathbf{A}}^{(h+1)}.
\end{equation}

The resulting orthonormal learned basis improves the conditioning of the next $\mathbf{A}$-update
without changing the current low-rank reconstruction.
\vspace{-2mm}
\subsubsection{Update for corruption component $\mathbf{S}$}
\label{subsubsec:update_S}

The corruption component is updated while holding $\mathbf{A}$, $\mathbf{B}$, and $\mathbf{Y}$
fixed. Because the $\ell_1$ penalty is non-differentiable, the closed-form solution to this
subproblem is obtained by setting the subgradient of the augmented Lagrangian with respect to
$\mathbf{S}$ to zero, $\mathbf{0}\in\partial_{\mathbf{S}}\mathcal{L}_{\nu^{(h)}}$. This leads to the
element-wise soft-thresholding update

\begin{equation}
    \label{eq:update_S_main}
    \mathbf{S}^{(h+1)} = \mathcal{S}_{1/\nu^{(h)}} \left( \mathbf{X} -
        \boldsymbol{\Psi}\mathbf{B}^{(h+1)}\mathbf{A}^{(h+1)} + \frac{1}{\nu^{(h)}}\mathbf{Y}^{(h)}
        \right),
\end{equation}

where $\mathcal{S}_{\tau}$ denotes the soft-thresholding operator with threshold $\tau$. This update
assigns an entry of the shifted residual to the corruption component only when its magnitude exceeds
$1/\nu^{(h)}$. The update order makes the initialization influential. The \(\mathbf{A}\)- and
\(\mathbf{B}\)-updates first construct \(\mathbf{L}^{(h+1)}\), after which soft thresholding is
applied to the shifted residual. During the first iteration,
\(\mathbf{S}^{(0)}=\mathbf{Y}^{(0)}=\mathbf{0}\), so equation~\eqref{eq:update_A_main} projects the
observed snapshot matrix onto the initial learned basis. Corrupted content represented in
\(\mathbf{L}^{(1)}\) is therefore removed from the residual before \(\mathbf{S}^{(1)}\) is computed
and is less likely to exceed the threshold \(1/\nu^{(0)}\). Later iterations can reassign this
content only if an update of \(\mathbf{A}\) or \(\mathbf{B}\) stops representing it, causing it to
reappear in the shifted residual. The alternating scheme does not guarantee such a transfer because
the Lagrange multiplier enforces \(\mathbf{X}=\mathbf{L}+\mathbf{S}\) without determining how
observed content is allocated between the two components. A smooth initial basis has limited overlap
with localized, rapidly varying corruption and therefore reduces the likelihood that such corruption
enters the early low-rank estimates. This initialization does not restrict the final learned basis
because subsequent updates of \(\mathbf{B}\) can incorporate finer library directions. Under the
continuation schedule defined in equation~\eqref{eq:nu_continuation}, the penalty parameter
$\nu^{(h)}$ increases as the optimization proceeds. The corresponding soft-thresholding level
$1/\nu^{(h)}$ therefore decreases over the iterations. During the early iterations, only relatively
large-magnitude entries of the shifted residual are assigned to $\mathbf{S}$. As the threshold
decreases, progressively lower-amplitude residual content can also enter the corruption component.
This continuation acts on the amplitude of the shifted residual rather than on the spatial size or
roughness of its structures. The spatial structures that can be represented in $\mathbf{L}$ are
instead determined by the current learned subspace. Consequently, decreasing the threshold can
recover weaker corruption that remains in the residual, but it does not by itself remove corruption
that has already been absorbed into the low-rank component.
\vspace{-2mm}
\subsubsection{Update for Lagrange multiplier $\mathbf{Y}$}
\label{subsubsec:update_Y}

After updating $\mathbf{A}$, $\mathbf{B}$, and $\mathbf{S}$, the Lagrange multiplier matrix is
updated using the current decomposition residual,

\begin{equation}
    \label{eq:update_Y_main}
    \mathbf{Y}^{(h+1)} = \mathbf{Y}^{(h)} + \nu^{(h)} \left( \mathbf{X} -
        \boldsymbol{\Psi}\mathbf{B}^{(h+1)}\mathbf{A}^{(h+1)} - \mathbf{S}^{(h+1)} \right).
\end{equation}

This update increases the multiplier in proportion to the remaining violation of the decomposition
constraint and drives the residual $\mathbf{X}-\boldsymbol{\Psi}\mathbf{B}\mathbf{A}-\mathbf{S}$
toward zero over the augmented Lagrangian iterations.
\vspace{-2mm}
\subsection{Convergence criterion}
\label{subsec:convergence_criterion}

Convergence is assessed using the decomposition residual and the relative change of the recovered
low-rank and corruption components. 
At iteration $h+1$, the relative decomposition residual is
defined as
\begin{equation}
    r_{\mathrm{residual}}^{(h+1)} = \frac{ \left\| \mathbf{X} - \mathbf{L}^{(h+1)} -
        \mathbf{S}^{(h+1)} \right\|_F }{ \max\left( \|\mathbf{X}\|_F, \delta \right) }, \qquad
        \mathbf{L}^{(h+1)} = \boldsymbol{\Psi} \mathbf{B}^{(h+1)} \mathbf{A}^{(h+1)},
    \label{eq:decomposition_residual}
\end{equation}
where $\delta>0$ is a numerical denominator safeguard. Since the low-rank and corruption components
are initialized as $\mathbf{L}^{(0)}=\mathbf{0}$ and $\mathbf{S}^{(0)}=\mathbf{0}$, direct
normalization by the previous iterate would be undefined during the first iteration. We therefore
define the combined relative change as
\begin{equation}
    r_{\mathrm{change}}^{(h+1)} = \max \left\{ \frac{ \left\| \mathbf{L}^{(h+1)} - \mathbf{L}^{(h)}
        \right\|_F }{ \max \left( \left\| \mathbf{L}^{(h)} \right\|_F, \delta \right) }, \frac{
        \left\| \mathbf{S}^{(h+1)} - \mathbf{S}^{(h)} \right\|_F }{ \max \left( \left\|
        \mathbf{S}^{(h)} \right\|_F, \delta \right) } \right\}.
    \label{eq:safeguarded_component_change}
\end{equation}
In the numerical implementation, $\delta$ is set to machine precision. This safeguard is active only
when the norm of the preceding iterate is zero or numerically negligible. A run is classified as
converged when
\begin{equation}
    r_{\mathrm{residual}}^{(h+1)} < \varepsilon_{\mathrm{residual}}, \qquad
        r_{\mathrm{change}}^{(h+1)} < \varepsilon_{\mathrm{change}}.
    \label{eq:convergence_conditions}
\end{equation}
These conditions require both satisfaction of the decomposition constraint and stabilization of the
recovered low-rank and corruption components. In the present analysis, we use
$\varepsilon_{\mathrm{residual}}=\varepsilon_{\mathrm{change}}=10^{-6}$. A tolerance-sensitivity
check showed that the reconstruction metrics changed only marginally when the tolerances
were varied over $10^{-4}$ to $10^{-8}$.
\vspace{-2mm}
\subsection{Final reconstruction}
\label{subsec:final_reconstruction}

After convergence, the final coefficient matrices are denoted by $\mathbf{A}^{*}$, $\mathbf{B}^{*}$,
and $\mathbf{S}^{*}$, and the final learned basis is
\begin{equation}
    \boldsymbol{\Phi}^{*} = \boldsymbol{\Psi}\mathbf{B}^{*}.
    \label{eq:final_learned_basis}
\end{equation}

The reconstructed low-rank snapshot matrix shown in figure~\ref{fig:method_overview} is computed as
\begin{equation}
    \mathbf{L}^{*} = \boldsymbol{\Phi}^{*}\mathbf{A}^{*} =
        \boldsymbol{\Psi}\mathbf{B}^{*}\mathbf{A}^{*}.
    \label{eq:final_low_rank_reconstruction}
\end{equation}
Each column of $\mathbf{L}^{*}$ corresponds to one recovered denoised snapshot. These columns are
reshaped back to the original spatial grid to obtain the reconstructed flow-field sequence. The
corruption matrix $\mathbf{S}^{*}$ contains the residual content assigned to the corruption
component, so the final decomposition satisfies the equality constraint to within the prescribed
convergence tolerance,
\begin{equation}
    \mathbf{X} \approx \mathbf{L}^{*} + \mathbf{S}^{*}.
    \label{eq:final_decomposition}
\end{equation}

\section{Results \& discussion}
\label{sec:results}

\subsection{Numerically simulated two-dimensional laminar post-stall NACA0012 wake}
\label{subsec:results_2d_naca}

We first consider the two-dimensional laminar post-stall NACA0012 wake at an angle of attack \(\alpha=40^\circ\) and a chord-based Reynolds number \(Re=100\).
The dataset is produced by direct numerical simulation~\cite{cliff1,cliff2,fukami2023grasping} over the domain
\((x,y)/c\in[-15,30]\times[-20,20]\),
with the leading edge of the airfoil at the origin.
The current simulation has been verified and validated compared to previous studies~\cite{ZFAT2023,kurtulus2015unsteady,liu2012numerical,di2018fluid}.
At the current condition, the wake exhibits periodic vortex shedding.
The snapshot matrix contains \(50\) uniformly sampled fields of the transverse velocity component \(v\), spanning approximately one shedding period.
The airfoil body and other invalid locations are masked, and the remaining data are normalized by the maximum absolute value of the uncorrupted dataset.

Because LLA-RPCA operates on a scalar snapshot matrix, the streamwise and transverse velocity components \(u\) and \(v\) are reconstructed separately.
The transverse component \(v\) is used for the primary reconstruction and modal comparisons.
The streamwise component \(u\) is additionally reconstructed for the calculation of spanwise vorticity from the reconstructed velocity field.
For LLA-RPCA, the prescribed modal capacity is \(r=5\), and the raw candidate library contains \(500\) trigonometric and \(500\) graph-Laplacian candidates.
The same corrupted matrix is supplied to RPCA and LLA-RPCA, while the uncorrupted sequence is used only for evaluation.
For the controlled corruption experiments, the reported reconstruction metrics are averaged over five independent corruption realizations.
In this example, we consider two different corruption styles; namely, 1, random distribution and 2. shear-concentrated distribution.
The second one enables examining how the current technique can be used for realistic scenarios in which PIV measurements might present high error for structures with high spatial gradients.

\subsubsection{Spatially random corruption}
\label{subsubsec:results_2d_naca_random}

For spatially random corruption, a fraction \(\eta\) of the valid space-time entries is selected uniformly at random without replacement and replaced by spikes of magnitude \(10\sigma\), where \(\sigma\) is the standard deviation of the valid entries in the normalized uncorrupted matrix.
Positive and negative spikes are assigned to approximately equal numbers of selected entries.
Because the corruption is imposed after normalization, the replacement values are not restricted to the normalized clean-reference range \([-1,1]\) and may exceed this range.

\begin{figure}
    \centering
    \includegraphics[width=0.95\linewidth]{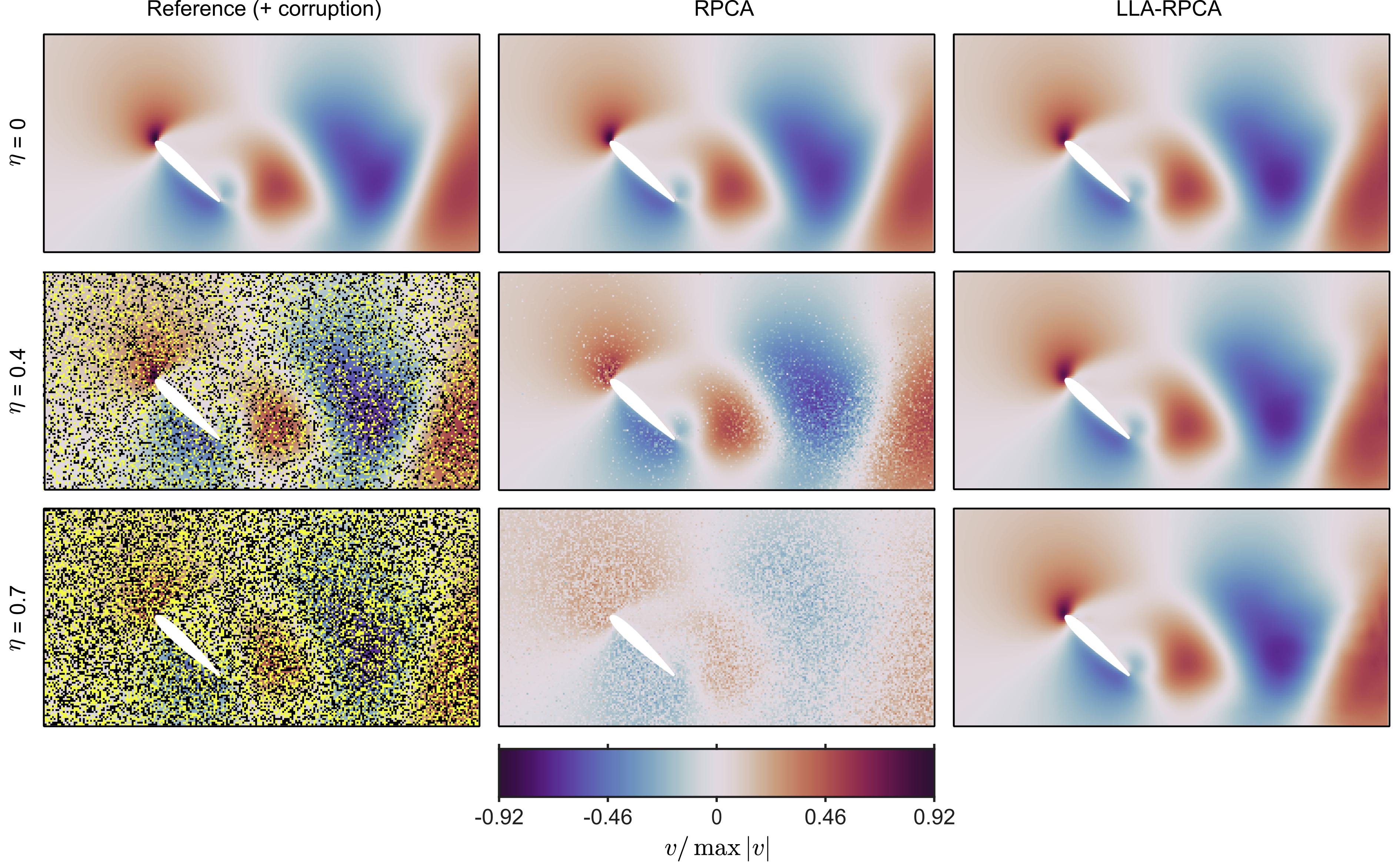}
    \caption{
    Reconstruction for the transverse velocity component \(v\) of the two-dimensional laminar post-stall NACA0012 wake under spatially random corruption at \(\eta=0\), \(\eta=0.4\), and \(\eta=0.7\), corresponding to \(0\%\), \(40\%\), and \(70\%\) imposed large-amplitude entrywise corruption.
    Values above and below the displayed color range are shown in yellow and black, respectively.
    }
    \label{fig:results_2d_naca_reconstruction_quality}
\end{figure}

Let us compare the reconstructed velocity fields at \(\eta=0\), \(\eta=0.4\), and \(\eta=0.7\) in figure~\ref{fig:results_2d_naca_reconstruction_quality}. 
As \(\eta\) increases, the corruption increasingly obscures the vortex-shedding pattern, although the dominant wake remains partly visible in the input at \(\eta=0.7\).
At \(\eta=0\), the RPCA reconstruction is visually similar to the reference field.
At \(\eta=0.4\), the magnitude of the positive and negative transverse velocity regions is reduced relative to the reference.
The attenuation becomes stronger at \(\eta=0.7\), while deviations from the reference spatial distribution also become more apparent.
This behavior is consistent with the physical-content-transfer failure mode described at the beginning of section~\ref{sec:method}.

For the imposed mixed-sign corruption, the attenuation also depends on the relative signs of the physical and corruption contributions.
In Equation~\eqref{eq:standard_rpca}, the \(\ell_1\) norm penalizes \(\mathbf{S}\) through the sum of its absolute entrywise values.
When physical content is assigned to an already corrupted entry in \(\mathbf{S}\), this penalty increases less when the physical and corruption contributions have opposite signs and can even decrease, making the transfer less costly or favorable from the sparse-penalty perspective.
As \(\eta\) increases, more physical entries overlap the corruption support, and therefore more locations are affected by this mechanism.
This results in increasingly widespread entrywise attenuation of the recovered velocity field toward zero.
At \(\eta=0.7\), this interpretation is supported by stronger attenuation at corrupted locations where the imposed corruption and the corresponding uncorrupted velocity have opposite signs.
LLA-RPCA instead suppresses the imposed corruption at both \(\eta=0.4\) and \(\eta=0.7\) while preserving the spatial distribution and magnitude of the positive and negative transverse velocity regions more closely.
The coherent wake can be represented by combinations of the prescribed candidate functions, whereas the spatially irregular mixed-sign corruption is less compatible with the candidate-library-constrained representation.
The improved reconstruction is therefore consistent with the two LLA-RPCA modifications introduced at the beginning of section~\ref{sec:method}: the spatial restriction limits the ability of the recovered component to represent the imposed corruption, while the explicit low-rank factorization avoids the nuclear-norm penalty associated with physical-content attenuation in standard RPCA.

To quantify these reconstruction differences over the complete corruption range, we use the total-field error \(e_{\mathrm{tot}}\), fluctuation error \(e_{\mathrm{fluc}}\), and structural similarity index error \(e_{\mathrm{SSIM}}\).
The total-field error measures the relative Frobenius-norm difference between the reconstructed field \(\mathbf{L}\) and the uncorrupted reference field \(\mathbf{X}_{\mathrm{ref}}\),
\begin{equation}
    e_{\mathrm{tot}}
    =
    \frac{
    \left\|\mathbf{L}-\mathbf{X}_{\mathrm{ref}}\right\|_F
    }{
    \left\|\mathbf{X}_{\mathrm{ref}}\right\|_F
    }.
\end{equation}
The fluctuation error evaluates only the time-dependent component of the reconstruction.
We first remove the temporal means from the reference and reconstructed fields,
\begin{equation}
    \mathbf{X}_{\mathrm{ref}}'
    =
    \mathbf{X}_{\mathrm{ref}}
    -
    \overline{\mathbf{X}}_{\mathrm{ref}},
    \qquad
    \mathbf{L}'
    =
    \mathbf{L}
    -
    \overline{\mathbf{L}},
\end{equation}
and then define the relative fluctuation error as
\begin{equation}
    e_{\mathrm{fluc}}
    =
    \frac{
    \left\|\mathbf{L}'-\mathbf{X}_{\mathrm{ref}}'\right\|_F
    }{
    \left\|\mathbf{X}_{\mathrm{ref}}'\right\|_F
    }.
\end{equation}
We further quantify the spatial agreement of individual snapshots using SSIM computed over local image regions~\cite{wang2004image}.
SSIM is less sensitive to pixel-wise errors caused by translational and rotational differences than a regular Frobenius-norm-based measure and has therefore been used for assessing fluid-flow reconstruction~\cite{anatharaman2023image,nakamura2022identifying}.
The SSIM value \(\chi\) is defined as
\begin{equation}
    \chi
    =
    l(i_x,i_y)c(i_x,i_y)s(i_x,i_y),
\end{equation}
where
\begin{equation}
    l(i_x,i_y)
    =
    \frac{2\mu_x\mu_y+C_1}{\mu_x^2+\mu_y^2+C_1},
    \qquad
    c(i_x,i_y)
    =
    \frac{2\sigma_x\sigma_y+C_2}{\sigma_x^2+\sigma_y^2+C_2},
    \qquad
    s(i_x,i_y)
    =
    \frac{\sigma_{xy}+C_3}{\sigma_x\sigma_y+C_3}.
\end{equation}
Here, \(i_x\) and \(i_y\) denote corresponding local regions in the reconstructed and reference snapshots.
The quantities \(\mu_x\) and \(\mu_y\) are their local means, \(\sigma_x\) and \(\sigma_y\) are their local standard deviations, and \(\sigma_{xy}\) is their local covariance.
The constants \(C_1\), \(C_2\), and \(C_3\) stabilize the division and are determined adaptively for each snapshot relative to its own signal range.
The resulting SSIM value is upper-bounded by one, representing identical fields.
For each snapshot \(j\), the SSIM value is computed between the reconstructed snapshot \(\mathbf{l}_j\) and the corresponding uncorrupted reference snapshot \(\mathbf{x}_{\mathrm{ref},j}\).
The reported SSIM error is
\begin{equation}
    e_{\mathrm{SSIM}}
    =
    1
    -
    \frac{1}{n}
    \sum_{j=1}^{n}
    \chi
    \left(
    \mathbf{l}_j,
    \mathbf{x}_{\mathrm{ref},j}
    \right).
\end{equation}
Lower values of \(e_{\mathrm{tot}}\), \(e_{\mathrm{fluc}}\), and \(e_{\mathrm{SSIM}}\) indicate better agreement with the uncorrupted reference.

\begin{figure}[t]
    \centering
    \includegraphics[width=0.95\linewidth]{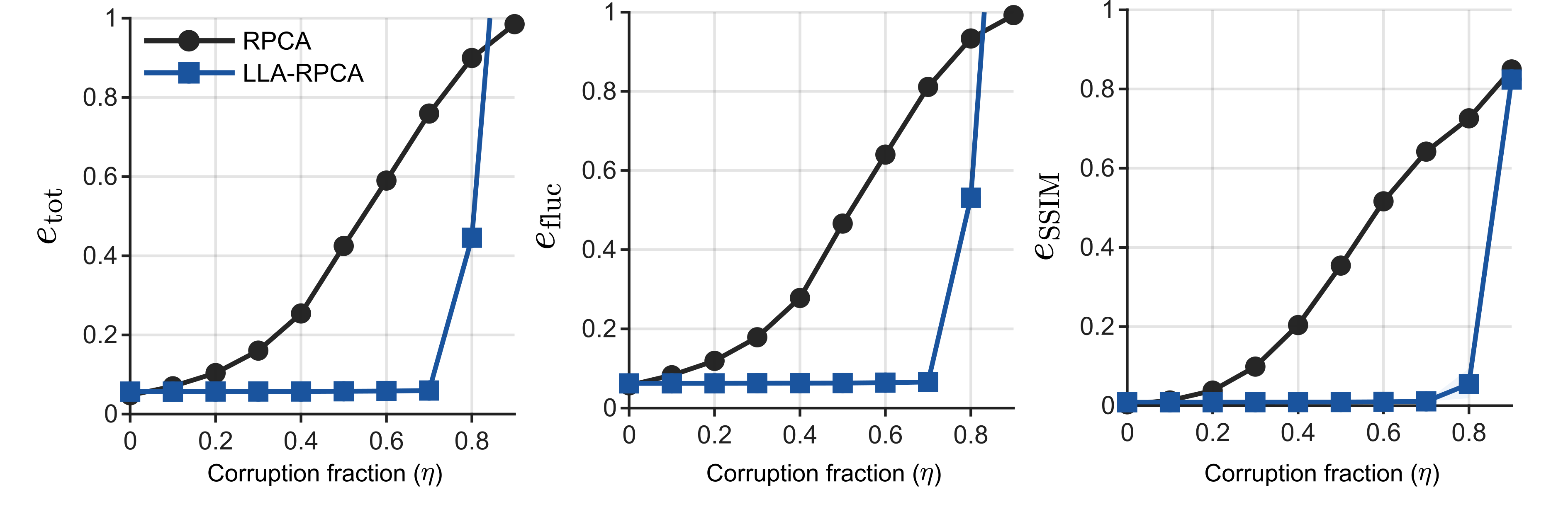}
    \caption{
    Error metrics for reconstruction of the normalized transverse velocity component \(v\) of the two-dimensional laminar post-stall NACA0012 wake under spatially random corruption as a function of \(\eta\).
    Each point denotes the mean over five independently generated corruption realizations at the corresponding value of \(\eta\).
    The spread over the five realizations is represented by a band, but the band is thinner than the plotted lines and is therefore not visible.
    }
    \label{fig:results_2d_naca_reconstruction_errors}
\end{figure}

The resulting error curves in figure~\ref{fig:results_2d_naca_reconstruction_errors} confirm the differences observed in the reconstructed fields.
The LLA-RPCA total-field and fluctuation errors remain nearly constant through \(\eta=0.7\), whereas the corresponding RPCA errors increase progressively.
The LLA-RPCA total-field error remains lower than the RPCA value through \(\eta=0.8\), and its SSIM error is lower across the intermediate corruption range.
The increasing RPCA errors are consistent with the progressive loss of physical velocity content visible in figure~\ref{fig:results_2d_naca_reconstruction_quality}.
At the largest corruption fractions, however, all three LLA-RPCA errors increase.
At these levels, only a small fraction of the entries in each snapshot remains unaffected by the imposed corruption.
The spatial basis is shared across the sequence, but the coefficient vector for each snapshot must still be inferred from the information available in that snapshot.
If the remaining reliable entries do not sufficiently sample the instantaneous wake, the corresponding coefficient vector becomes poorly constrained.
Because the temporal coefficients are not regularized across neighboring snapshots, this can lead to intermittent snapshot-level reconstruction failure.
The sharp increase in \(e_{\mathrm{tot}}\) and \(e_{\mathrm{fluc}}\) at the largest \(\eta\) therefore reflects the onset of this failure regime.

To determine whether the differences in velocity reconstruction also affect a subsequently derived flow quantity, we calculate spanwise vorticity at \(\eta=0.7\).
The streamwise and transverse velocity components \(u\) and \(v\) are reconstructed separately using the same corruption realization, after which the vorticity $\omega$
is calculated from the reconstructed velocity fields.
The resulting fields are compared with the vorticity calculated in the same manner from the uncorrupted reference velocities.

\begin{figure}[!b]
    \centering
    \includegraphics[width=0.95\linewidth]{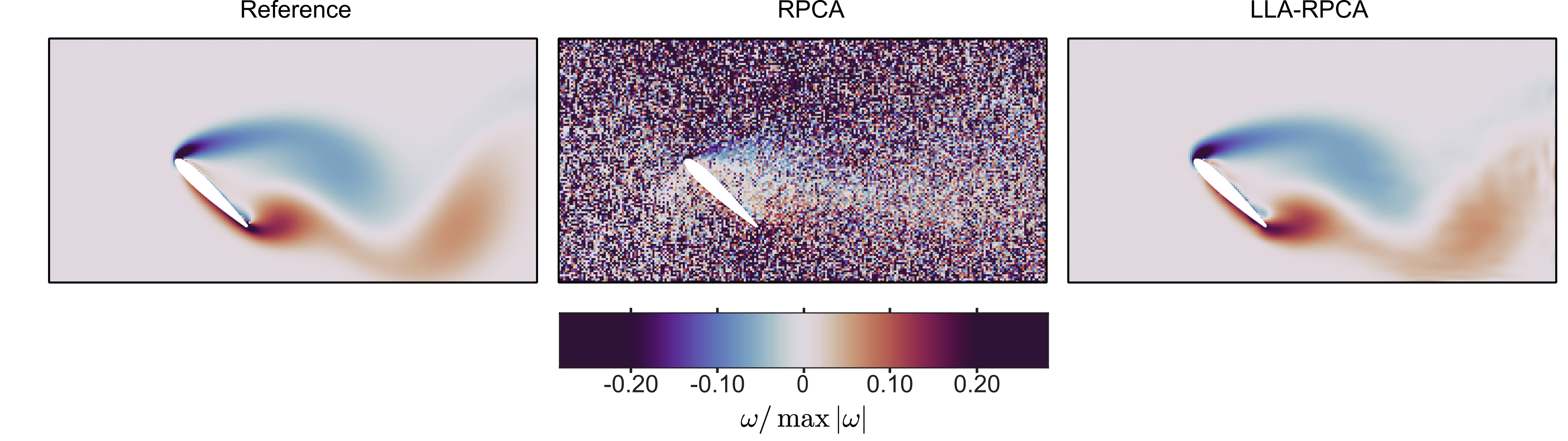}
    \caption{
    Spanwise vorticity fields computed from the uncorrupted reference velocities and from the RPCA- and LLA-RPCA-reconstructed velocity fields for the two-dimensional laminar post-stall NACA0012 wake at $\eta=0.7$, corresponding to 70\% imposed large-amplitude entrywise corruption.
    }
    \label{fig:results_2d_naca_vorticity}

    \vspace{6mm}

    \includegraphics[width=0.95\linewidth]{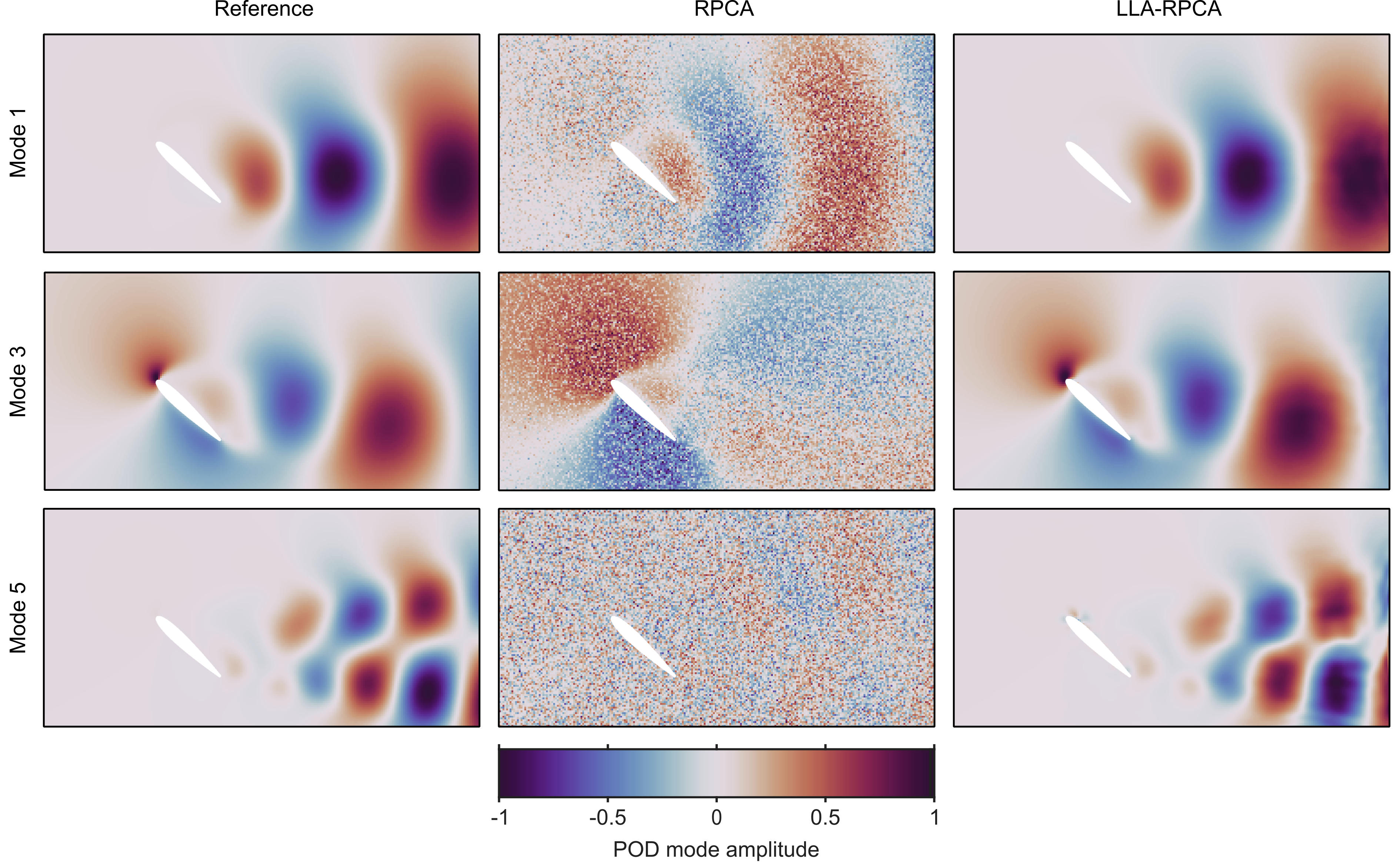}
    \caption{
    POD mode comparison for the transverse velocity sequence of the two-dimensional laminar post-stall NACA0012 wake under spatially random corruption at \(\eta=0.7\), corresponding to \(70\%\) imposed large-amplitude entrywise corruption.
    The modes are computed from the uncorrupted reference sequence and from the RPCA and LLA-RPCA reconstructions.
    }
    \label{fig:results_2d_naca_modal}
\end{figure}

Let us present in figure~\ref{fig:results_2d_naca_vorticity} the vorticity fields derived from the two reconstructions.
For RPCA, the dominant positive and negative vorticity regions are almost completely obscured, and the field is substantially more corrupted than the corresponding reconstructed velocity field.
This follows from the sensitivity of spatial differentiation to local velocity errors.
Spatially varying reconstruction errors in either velocity component can therefore be amplified by the derivative operation and can influence neighboring vorticity estimates.
The vorticity calculated from the LLA-RPCA reconstruction remains much closer to the uncorrupted reference.
The spatial distribution and magnitude of the dominant signed vorticity regions are retained, while the far field does not develop the widespread spurious structures observed for RPCA.

\begin{figure}[t]
    \centering
    \includegraphics[width=0.95\linewidth]{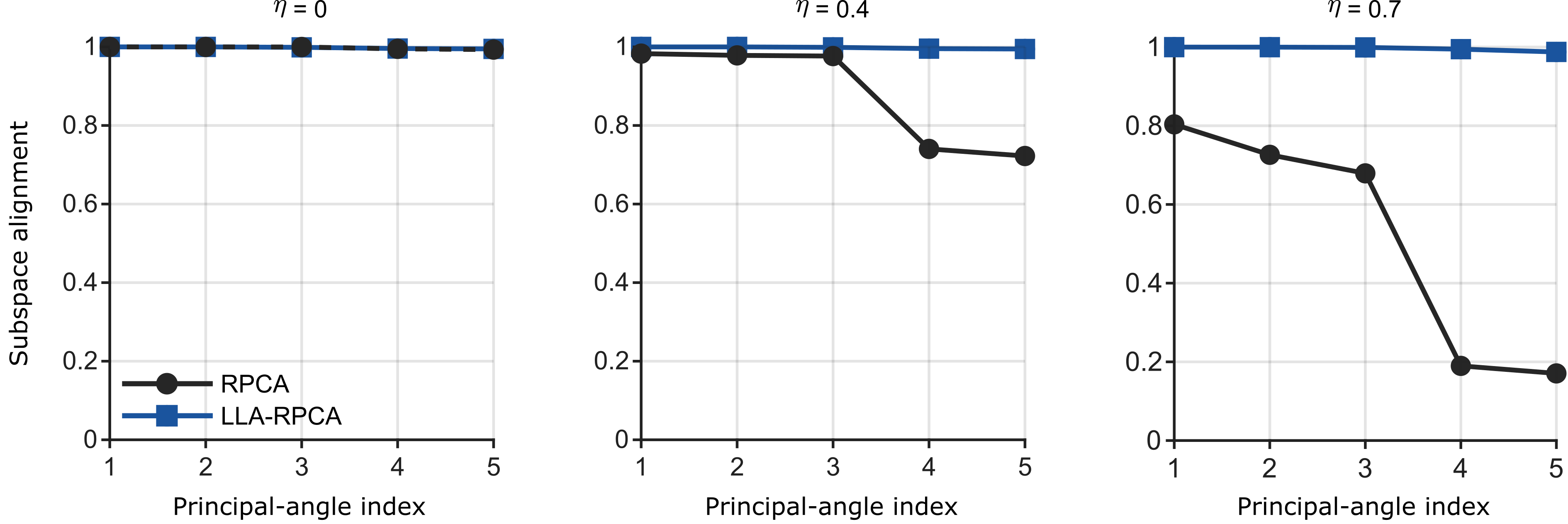}
    \caption{
    POD-subspace alignment for the transverse velocity sequence of the two-dimensional laminar post-stall NACA0012 wake under spatially random corruption at \(\eta=0\), \(\eta=0.4\), and \(\eta=0.7\), corresponding to \(0\%\), \(40\%\), and \(70\%\) imposed large-amplitude entrywise corruption.
    }
    \label{fig:results_2d_naca_subspace_alignment}
    \vspace{-3mm}
\end{figure}

We also examine whether the reconstruction differences propagate into the POD representation of the wake.
POD is applied to the complete RPCA and LLA-RPCA low-rank reconstructions, and the resulting modes are compared with those of the uncorrupted reference matrix, as depicted in figure~\ref{fig:results_2d_naca_modal}.
Here, we consider a fraction at \(\eta=0.7\).
The RPCA modes contain visible residual contamination, which becomes more pronounced in the higher displayed modes, and their spatial distributions increasingly deviate from those of the corresponding reference modes.
Although attenuation is the dominant error in the instantaneous RPCA reconstructions, lower-amplitude residual errors remain in the low-rank matrix and are separated by POD into individual modal directions.
The LLA-RPCA modes more closely preserve the spatial distribution and sign structure of the corresponding reference modes.
To quantify this modal agreement, POD subspace recovery is evaluated using principal angles between the reference and reconstructed POD subspaces.
For a single mode, this reduces to the angle between two vectors, while for multiple modes the principal angles identify the closest corresponding directions within the two subspaces.
Only directions associated with numerically nonzero singular values are included.
Let \(r_{\mathrm{ref}}\) and \(r_{\mathrm{rec}}\) denote the ranks of the reference and reconstructed snapshot matrices.
For a requested comparison dimension \(d_{\mathrm{requested}}\), the number of compared directions is
\begin{equation}
d
=
\min
\left(
d_{\mathrm{requested}},
r_{\mathrm{ref}},
r_{\mathrm{rec}}
\right).
\label{eq:subspace_comparison_dimension}
\end{equation}
The first \(d\) supported POD directions are collected as
\begin{equation}
\mathbf{U}_{\mathrm{ref},d}
=
[
\mathbf{u}_{\mathrm{ref},1},
\dots,
\mathbf{u}_{\mathrm{ref},d}
],
\qquad
\mathbf{U}_{\mathrm{rec},d}
=
[
\mathbf{u}_{\mathrm{rec},1},
\dots,
\mathbf{u}_{\mathrm{rec},d}
].
\end{equation}
The singular values of
\(\mathbf{U}_{\mathrm{ref},d}^{T}\mathbf{U}_{\mathrm{rec},d}\)
equal the cosines of the principal angles.
Alignment values close to one indicate strong subspace agreement, whereas values close to zero indicate poor alignment.
When the rank is lower than \(d_{\mathrm{requested}}\), no values are reported beyond the available rank because the corresponding singular vectors are not supported by nonzero singular values and do not represent physically meaningful POD directions.
Their omission therefore indicates rank collapse rather than zero alignment.
The alignment values in figure~\ref{fig:results_2d_naca_subspace_alignment} show how POD-subspace recovery changes at \(\eta=0\), \(\eta=0.4\), and \(\eta=0.7\).
At \(\eta=0\), both methods remain nearly perfectly aligned with the reference subspace across all five principal angles.
At \(\eta=0.4\), LLA-RPCA remains closely aligned throughout, whereas RPCA shows reduced agreement in the less-aligned subspace directions.
At \(\eta=0.7\), this separation becomes pronounced, with LLA-RPCA retaining high alignment while RPCA shows substantial disagreement with the reference subspace.

\subsubsection{Shear-based corruption}
\label{subsubsec:results_2d_naca_shear}

To examine whether the reconstruction behavior changes when the corruption is spatially concentrated in flow regions that are more prone to PIV errors, we additionally evaluate the proposed technique under a shear-based corruption.
In this case, corrupted locations are preferentially concentrated in regions of high shear rather than uniformly over the fluid domain.
This represents the tendency of unreliable PIV measurements to occur preferentially in flow regions where strong velocity gradients make the particle-image correlation more challenging.
The comparison is performed at \(\eta=0.4\), while the imposed corruption magnitude and mixed-sign replacement values remain unchanged.

\begin{figure}[!b]
    \centering
    \includegraphics[width=0.95\linewidth]{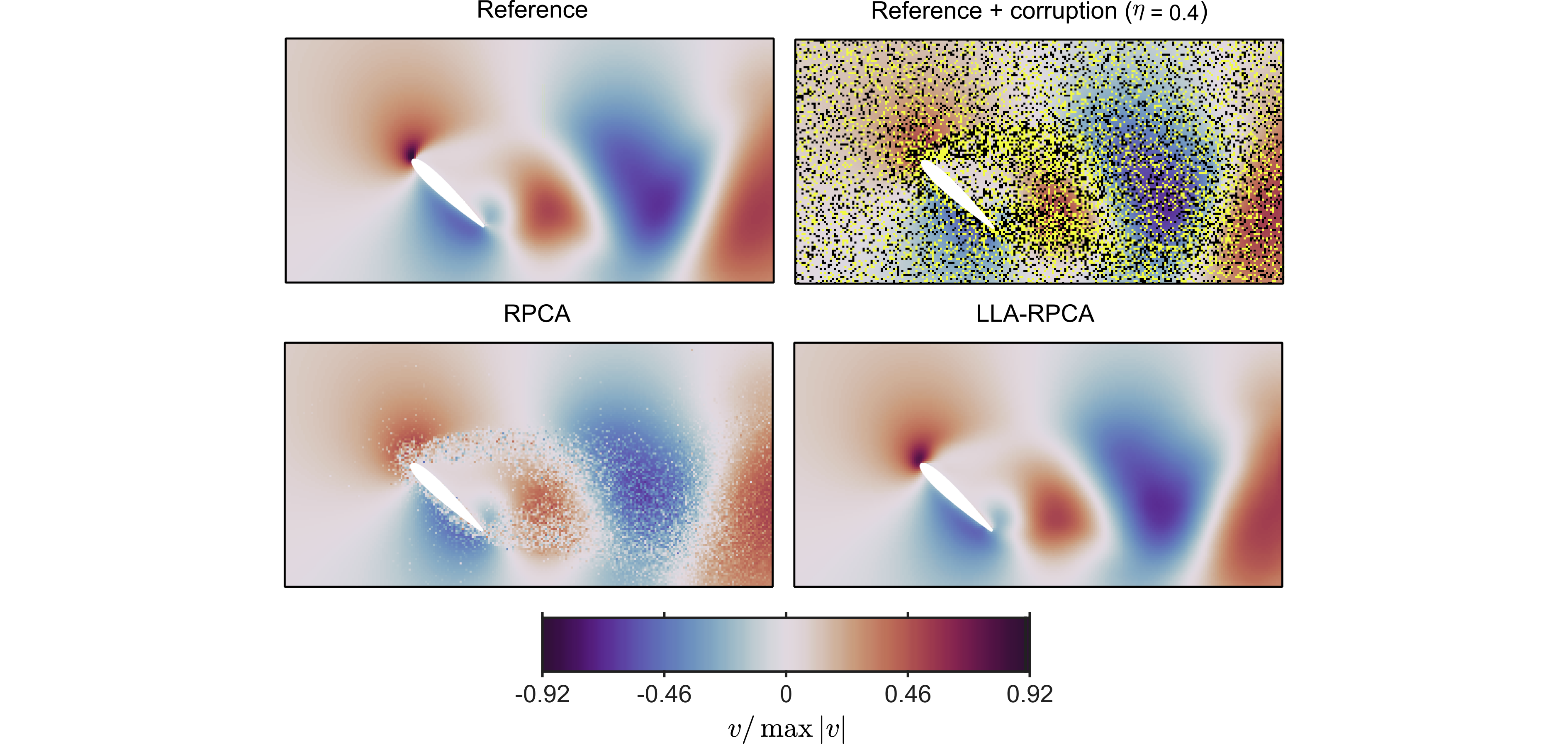}
    \caption{
    Reconstruction results for the normalized transverse velocity component \(v\) of the two-dimensional laminar post-stall NACA0012 wake under shear-based corruption at \(\eta=0.4\), corresponding to \(40\%\) imposed large-amplitude entrywise corruption.
    Values above and below the displayed color range are shown in yellow and black, respectively.
    }
    \label{fig:results_2d_naca_shear_reconstruction}

    \vspace{10mm}

    \includegraphics[width=0.95\linewidth]{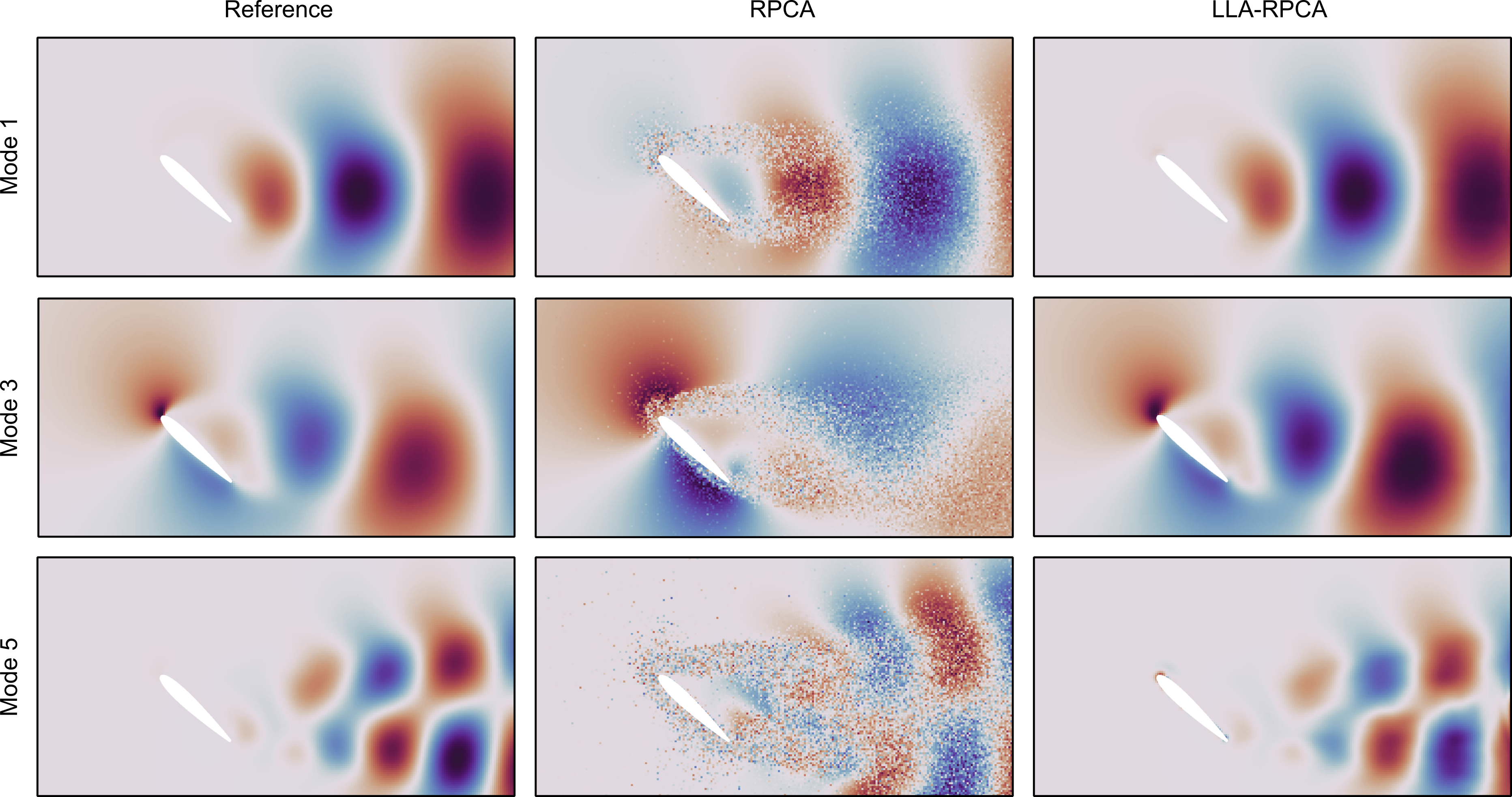}
    \caption{
    POD mode comparison for the transverse velocity sequence of the two-dimensional laminar post-stall NACA0012 wake under shear-based corruption at \(\eta=0.4\), corresponding to \(40\%\) imposed large-amplitude entrywise corruption.
    }
    \label{fig:results_2d_naca_shear_modal}
\end{figure}

We show in figure~\ref{fig:results_2d_naca_shear_reconstruction} that the imposed corruption is concentrated primarily along the high-shear regions extending from the leading and trailing edges, while the surrounding far field is affected less strongly.
The RPCA reconstruction retains the overall spatial distribution of the positive and negative transverse velocity regions.
However, their magnitude is strongly attenuated in the high-shear regions and smaller pixel-scale residual errors remain visible.
The reconstruction is more accurate in the weakly corrupted far field.
The behavior therefore resembles the spatially random case at different local corruption levels: RPCA performs well where relatively few entries are corrupted but loses physical content where the corruption is locally concentrated.

LLA-RPCA instead closely reproduces the uncorrupted reference field throughout the domain.
The spatial distribution and magnitude of the positive and negative transverse velocity regions are retained in both the weakly and strongly corrupted regions, and no residual corruption is visible.
Although the corruption support is concentrated in regions of high shear, the imposed values remain spatially irregular at the corrupted entries and are less compatible with the candidate-library-constrained representation than the coherent velocity structures.
This is consistent with the spatial-restriction mechanism introduced at the beginning of section~\ref{sec:method}, showing that it remains effective when the local corruption fraction varies strongly across the domain.


To determine whether these spatially concentrated reconstruction errors also affect the recovered modal structures, we compare the corresponding POD modes in figure~\ref{fig:results_2d_naca_shear_modal}.
The RPCA modes contain pronounced deviations concentrated in the same regions preferentially affected by the imposed corruption.
The loss of velocity content in these regions also alters the spatial structure of the recovered POD modes.
The LLA-RPCA modes reproduce the reference modal structures much more closely, with the spatial distribution and sign structure of the dominant modal regions retained and only small localized deviations toward the downstream end of the wake in the higher displayed mode.
The shear-based case therefore shows that the improved reconstruction and modal recovery of LLA-RPCA are retained when the corruption is preferentially concentrated in physically relevant high-shear regions rather than distributed uniformly across the flow field.

\begin{figure}
    \centering
    \includegraphics[width=0.95\linewidth]{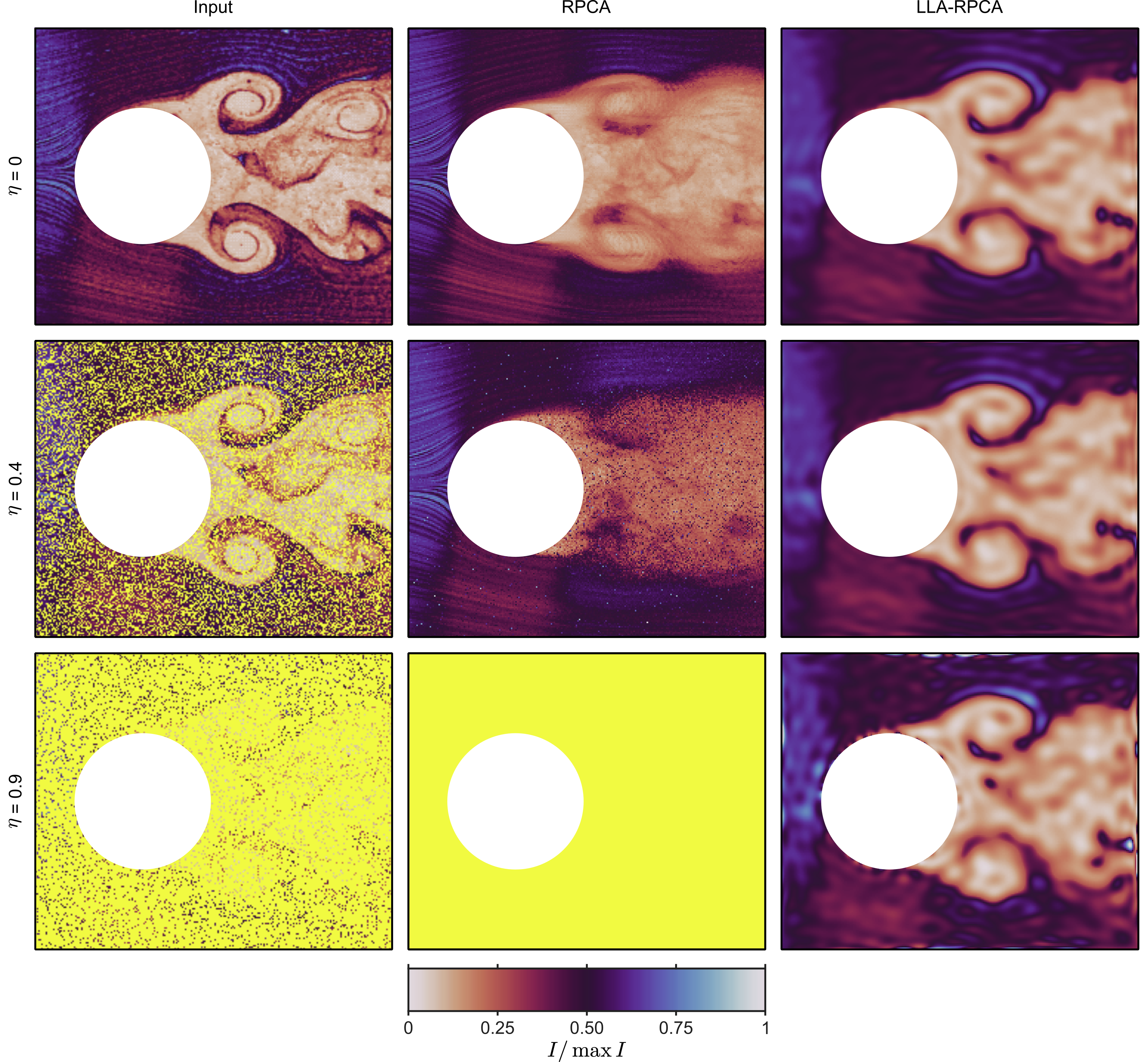}
    \caption{
    Reconstruction results for the oscillating-cylinder wake at \(\eta=0\), \(\eta=0.4\), and \(\eta=0.9\), corresponding to \(0\%\), \(40\%\), and \(90\%\) imposed large-amplitude entrywise corruption.
    Values above the displayed color range are shown in yellow.
    }
    \label{fig:results_cylinder_reconstruction_quality}
\end{figure}

\subsection{Video data of oscillating-cylinder wake}
\label{subsec:results_cylinder}

\begin{figure}[!b]
    \centering
    \includegraphics[width=0.95\linewidth]{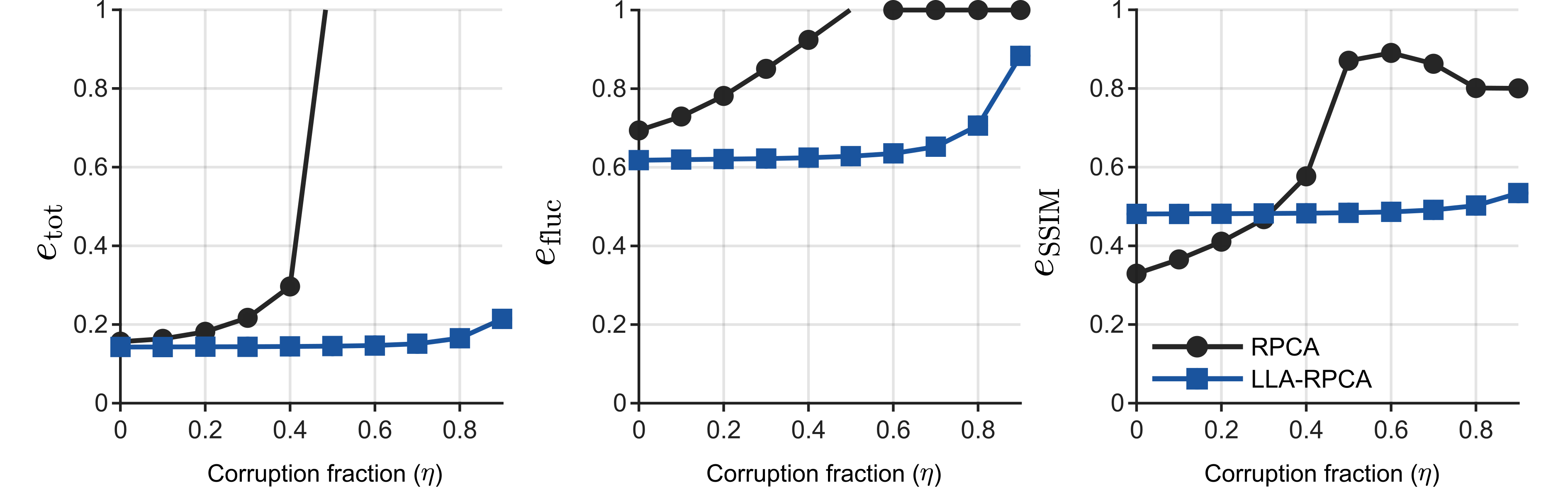}
    \caption{
    Error metrics for the oscillating-cylinder wake as a function of \(\eta\).
    Each point denotes the mean over five independently generated corruption realizations at the corresponding value of \(\eta\).
    The spread over the five realizations is represented by a band, but the band is thinner than the plotted lines and is therefore not visible.
    }
    \label{fig:results_cylinder_reconstruction_errors}
\end{figure}

We next consider a flow-visualization video from the American Physical Society Gallery of Fluid Motion entry by Boersma et al.~\cite{boersma2021vortexarms}.
The video shows symmetric and alternating shedding behind a cylinder oscillating in the streamwise direction.
This data set has also been used to assess data-driven reduced-order modeling methods~\cite{tomasetto2025reduced}.
Although the cylinder motion is periodic, the observed wake does not repeat exactly between successive cycles and exhibits quasi-periodic variations in the streakline and vortex patterns.

\begin{figure}[t]
    \centering
    \includegraphics[width=0.95\linewidth]{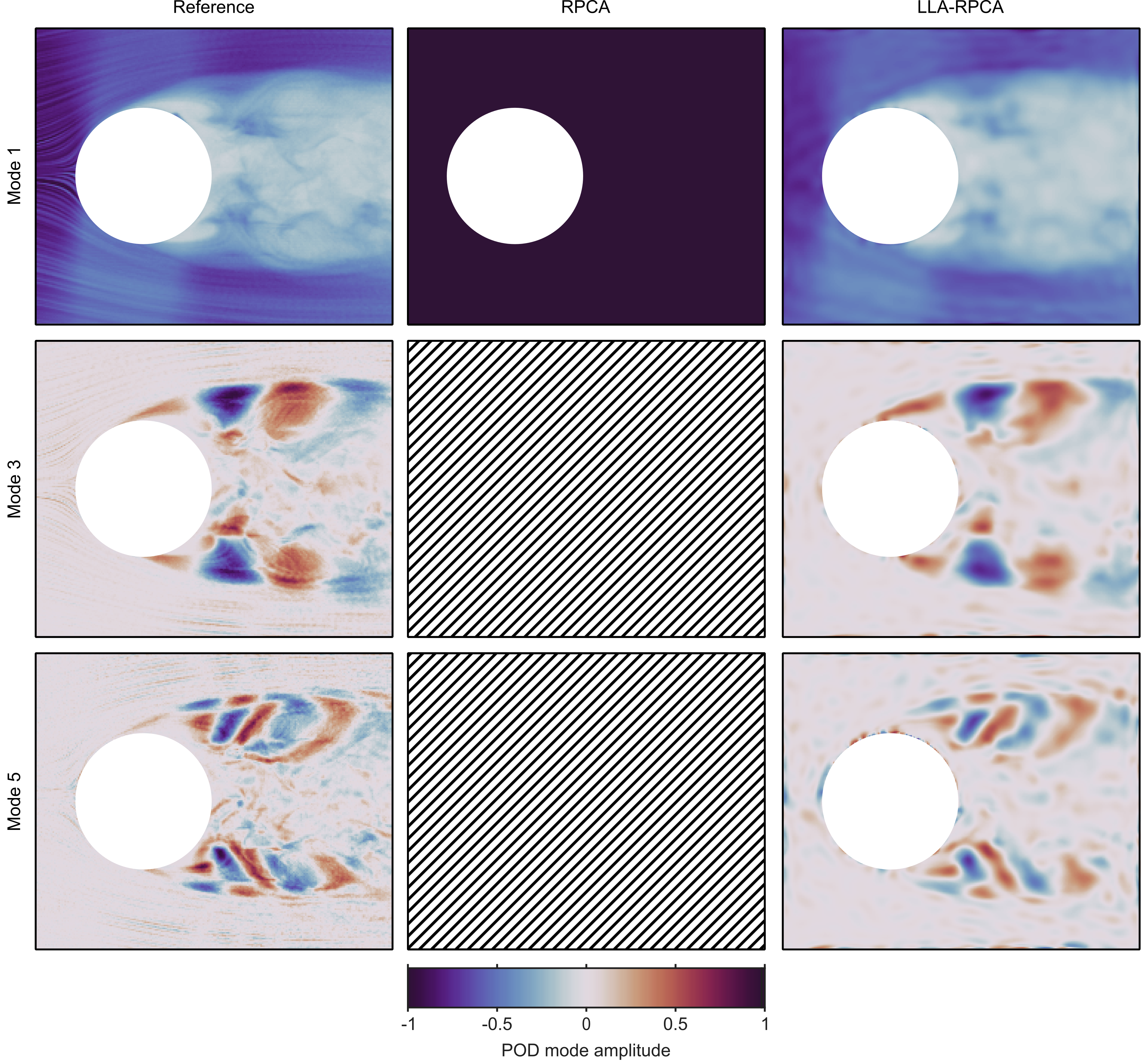}
    \caption{
    POD comparison for the oscillating-cylinder wake at \(\eta=0.9\), corresponding to \(90\%\) imposed large-amplitude entrywise corruption.
    The displayed RPCA low-rank reconstruction has rank one, so only its first POD direction is supported by a nonzero singular value.
    The diagonally hatched RPCA panels for the displayed third and fifth directions indicate that these higher-order modes are absent and are not physically interpretable POD modes.
    }
    \label{fig:results_cylinder_modal}
\end{figure}

Unlike the NACA0012 data, the snapshots contain experimental image intensities rather than a computed flow variable.
Each frame is converted to grayscale and registered in a cylinder-fixed coordinate system.
The snapshot matrix contains \(56\) uniformly sampled frames spanning approximately one cylinder-oscillation period.
The cylinder and other invalid regions are masked, and the remaining intensities are normalized by the maximum intensity of the uncorrupted reference sequence.
For each \(\eta\), a fraction \(\eta\) of the valid space-time entries is selected uniformly at random without replacement and replaced by positive spikes of magnitude \(10\sigma\).
Because the corruption is imposed after normalization, the replacement values are not restricted to the normalized clean-reference range \([0,1]\) and may exceed this range.
The corrupted intensity matrix is then transformed to a signed model-space representation using a continuous piecewise-affine mapping because the prescribed library contains alternating-sign basis functions.
The inverse mapping is applied after reconstruction.
The imposed corruption remains positive at every selected entry under this transformation, so both methods receive positive-only corruption in model space.
RPCA and LLA-RPCA are supplied with the same transformed corrupted matrix, and their reconstructed low-rank components are mapped back to the grayscale-intensity domain before comparison with the uncorrupted reference sequence.
The reported metrics are averaged over five independent corruption realizations.
For LLA-RPCA, \(r=17\), and the raw candidate library contains \(500\) trigonometric and \(500\) graph-Laplacian candidates.

The reconstructed image sequences at \(\eta=0\), \(\eta=0.4\), and \(\eta=0.9\) are compared in figure~\ref{fig:results_cylinder_reconstruction_quality}.
The imposed corruption progressively obscures the instantaneous wake in the input images.
At \(\eta=0\), RPCA retains the broad wake envelope but does not accurately preserve the instantaneous vortex positions and intensity distribution.
At \(\eta=0.4\), the vortex-arm structure is lost and the reconstruction is dominated by a contaminated mean-wake pattern.
At \(\eta=0.9\), the RPCA reconstruction collapses to an almost spatially and temporally uniform field whose intensity is numerically equal to the imposed positive corruption level of \(10\sigma\).
Its low-rank matrix has rank one.
At this corruption fraction, \(90\%\) of the observed entries have been replaced by the same positive value, so the observed matrix is dominated by an approximately constant rank-one pattern.
Standard RPCA can therefore favor a decomposition in which this dominant repeated-value pattern is assigned to the low-rank component \(\mathbf{L}\), while the minority of entries that retain the original image intensities are represented as deviations in the sparse component \(\mathbf{S}\).
The intended roles of the two components are thus effectively reversed.

The RPCA behavior therefore reflects both failure modes described at the beginning of section~\ref{sec:method}.
The loss of instantaneous wake structure at lower corruption fractions is consistent with physical-content transfer, whereas the collapse to the imposed corruption level at \(\eta=0.9\) demonstrates substantial contamination of the recovered low-rank component.
At this corruption level, only one numerically nonzero singular direction remains, and the resulting reconstruction no longer represents the time-dependent wake.
Because standard RPCA does not prescribe the retained modal capacity directly, the rank-one reconstruction also leaves insufficient spatial directions to represent the changing wake.
This particular contamination route is unavailable to LLA-RPCA because the cylinder library excludes a constant spatial candidate, so the dominant constant replacement pattern cannot be represented directly by the recovered low-rank field.
LLA-RPCA instead retains the wake envelope and instantaneous vortex positions throughout the range, with limited visible degradation at \(\eta=0.9\).
Its prescribed modal capacity \(r=17\) provides multiple spatial directions for representing the changing wake, while the candidate-library restriction limits the ability of the learned basis to reproduce the remaining spatially irregular corruption.
The LLA-RPCA images remain smoother than the reference, particularly in the background, because the finite library inadequately resolves the thin, predominantly horizontal seeding-related streaklines.
This illustrates a limitation of the candidate-library approach: spatial content that is not sufficiently represented by the candidate set can also be removed or smoothed.
In this case, the loss mainly concerns fine image texture, while the principal wake evolution remains recognizable.

The reconstruction differences across the full corruption range are quantified by the error curves in figure~\ref{fig:results_cylinder_reconstruction_errors}.
LLA-RPCA has lower total-field and fluctuation errors than RPCA at every tested corruption fraction.
Both LLA-RPCA errors change only slightly over most of the tested range and increase toward the largest corruption fractions, whereas the corresponding RPCA errors increase more strongly as the reconstructed wake progressively deteriorates.
At lower and intermediate corruption fractions, this deterioration includes attenuation and loss of instantaneous wake structure, while at \(\eta=0.9\) the rank-one collapse to the imposed positive corruption level demonstrates substantial contamination of the recovered low-rank component.
The SSIM comparison differs at low corruption fractions.
For \(\eta\leq0.3\), RPCA has the lower SSIM error because the LLA-RPCA reconstruction smooths the thin seeding-related streaklines and repeated local background structures that contribute to the localized SSIM calculation.
Loss of this fine texture can therefore increase the SSIM error even when the larger-scale wake structure is reconstructed more accurately.
From \(\eta=0.4\) onward, the progressive loss of the principal wake structure in the RPCA reconstruction becomes dominant and LLA-RPCA has the lower SSIM error.

To examine whether the loss of instantaneous wake structure is also reflected in the recovered modal representation, we apply the same POD procedure used for the NACA0012 case, as presented in figure~\ref{fig:results_cylinder_modal}.
We consider the condition of \(\eta=0.9\), where only \(10\%\) of the input entries remain uncorrupted.
The LLA-RPCA modes retain the dominant image structure and alternating wake patterns of the corresponding reference modes.
They are smoother than the reference modes because some fine-scale image content is not represented by the finite candidate library, but the principal wake structures and sign distributions remain recognizable.

The RPCA low-rank matrix, in contrast, has rank one at this corruption level.
Only its first POD direction is associated with a nonzero singular value.
This remaining direction corresponds to the nearly spatially uniform reconstruction rather than the instantaneous wake structure.
The hatched third- and fifth-mode panels in figure~\ref{fig:results_cylinder_modal} denote absent modes rather than weakly recovered physical modes.
The reconstruction has collapsed to one spatial direction and cannot represent the time-dependent wake evolution.

\begin{figure}[t]
    \centering
    \includegraphics[width=0.95\linewidth]{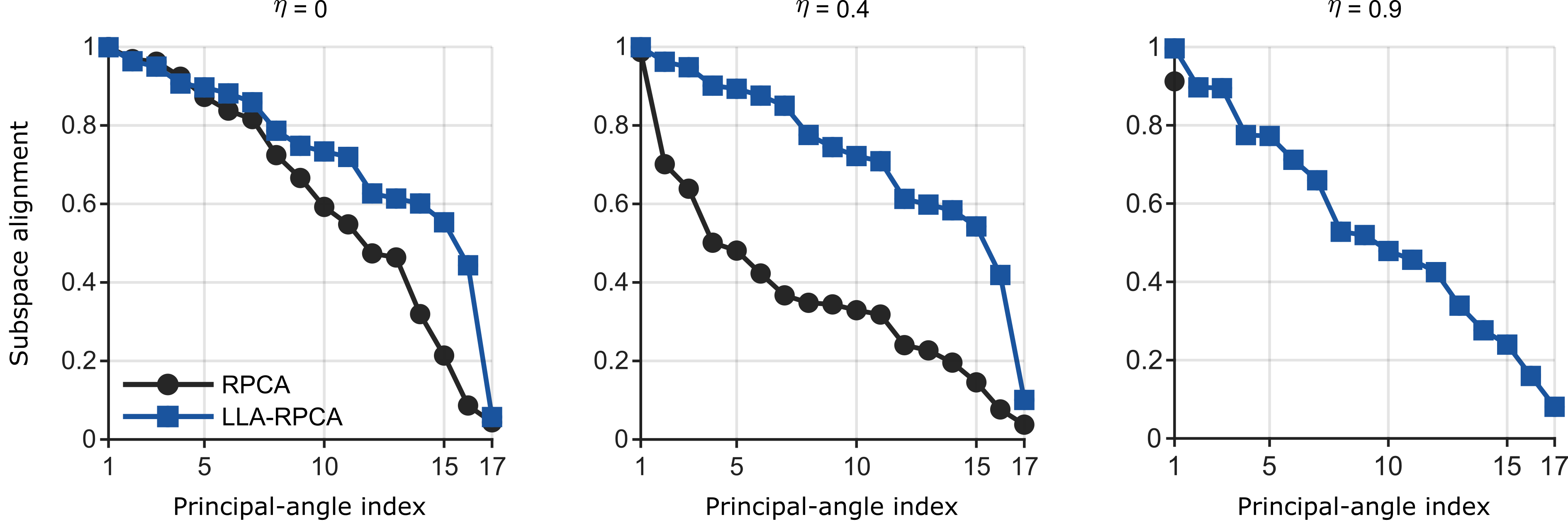}
    \caption{
    POD-subspace alignment for the oscillating-cylinder wake at \(\eta=0\), \(\eta=0.4\), and \(\eta=0.9\), corresponding to \(0\%\), \(40\%\), and \(90\%\) imposed large-amplitude entrywise corruption.
    At \(\eta=0.9\), the RPCA low-rank reconstruction has rank one, so only its first subspace-alignment value is defined and shown.
    }
    \label{fig:results_cylinder_subspace_alignment}
\end{figure}

The corresponding principal-angle alignment in figure~\ref{fig:results_cylinder_subspace_alignment} shows increasing separation between the reconstructed and reference POD subspaces as the corruption fraction increases.
At \(\eta=0\), LLA-RPCA shows higher alignment than RPCA over most of the principal-angle spectrum.
At \(\eta=0.4\), the difference becomes substantially larger, with LLA-RPCA retaining considerably stronger subspace agreement.
At \(\eta=0.9\), LLA-RPCA still retains a multidimensional subspace with substantial agreement in its most closely aligned directions.
For RPCA, only one principal angle is defined because the reconstructed matrix has collapsed to rank one.

\subsection{Experimentally measured transverse extreme gust encounter}
\label{subsec:results_flat_plate}

Finally, we consider time-resolved PIV measurements of a flat plate undergoing a transverse extreme gust encounter.
The plate is held at \(\alpha=0^\circ\) and towed through a transverse jet at a chord-based Reynolds number \(Re=20{,}000\).
The gust ratio is
\(G=V_{\rm max}/u_\infty=1.5\),
where \(V_{\rm max}\) is the peak gust velocity and \(u_\infty\) is the freestream speed.
This ratio characterizes the disturbance strength relative to the forward velocity, and conditions with \(G\geq1\) may occur in urban canyons, mountainous terrain, and ship wakes~\cite{jones2022physics,taira2026extreme,fukami2025extreme}.
The large aerodynamic response over a short duration makes this a strongly unsteady transient case.

The full dataset contains \(1284\) velocity fields over \(0\leq t\leq6.33\).
Here, the analysis is restricted to the interval \(0.50\leq t\leq2.97\), which contains the strongest transient behavior during the gust encounter.
From this interval, \(100\) uniformly spaced snapshots are selected for each velocity component to form the snapshot matrices used for the analysis.
The plate is masked in all velocity fields.
The experimental setup and data curation are described by Biler et al.~\cite{biler2021experimental}, and the data are made available by Towne et al.~\cite{towne2023database}.
No synthetic corruption is imposed, and no uncorrupted reference sequence is available.

\begin{figure}[t]
    \centering
    \includegraphics[width=0.95\linewidth]{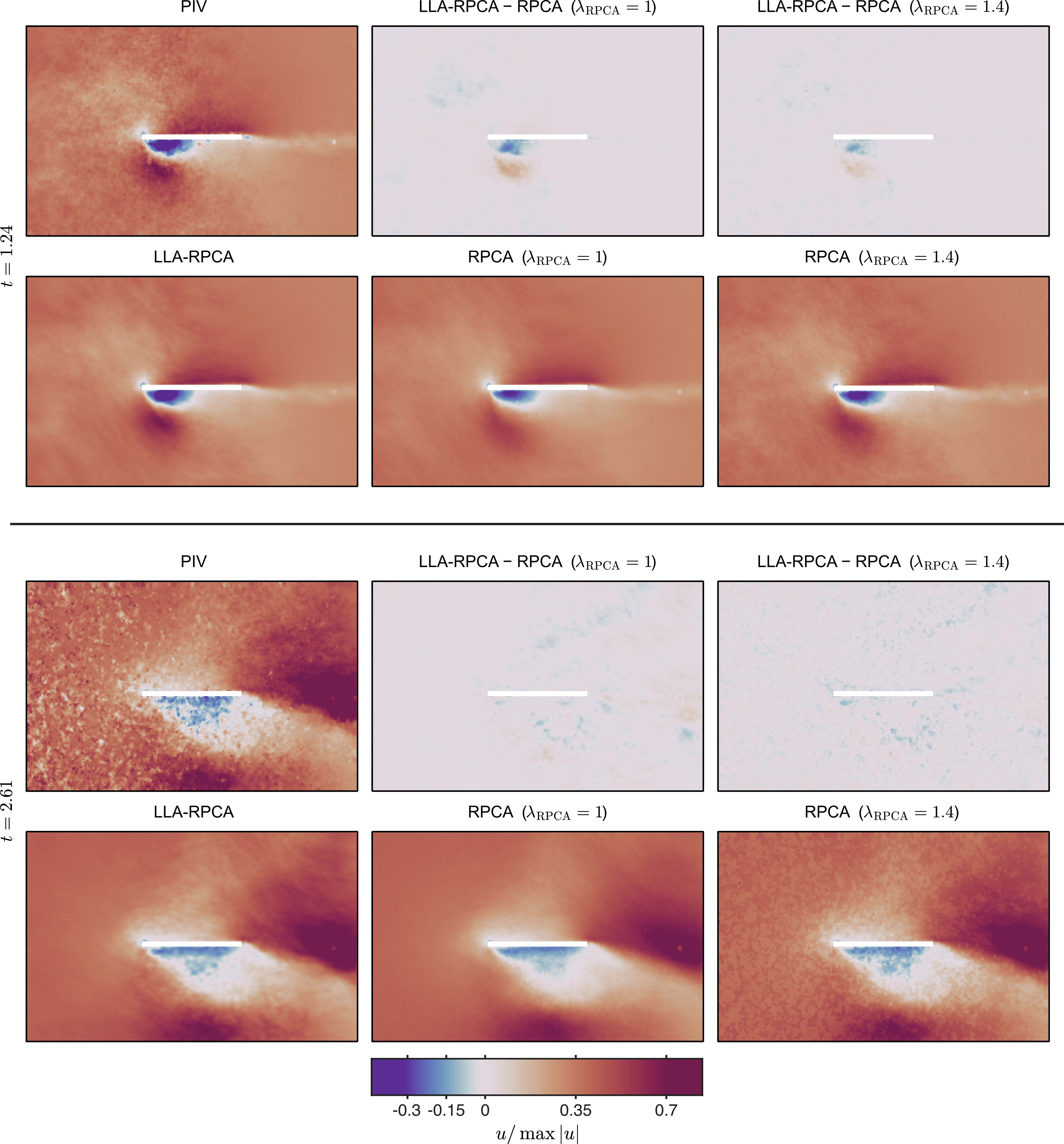}
    \caption{
    Reconstruction results for the streamwise velocity during the experimentally measured transverse extreme gust encounter at \(t=1.24\) and \(t=2.61\).
    For each snapshot, shown are the measured PIV field, the LLA-RPCA reconstruction, the RPCA reconstructions with \(\lambda_{\mathrm{RPCA}}=1\) and \(1.4\), and the corresponding differences between the LLA-RPCA and RPCA reconstructions.
    }
    \label{fig:Plate_u}
\end{figure}

\begin{figure}[t]
    \centering
    \includegraphics[width=0.95\linewidth]{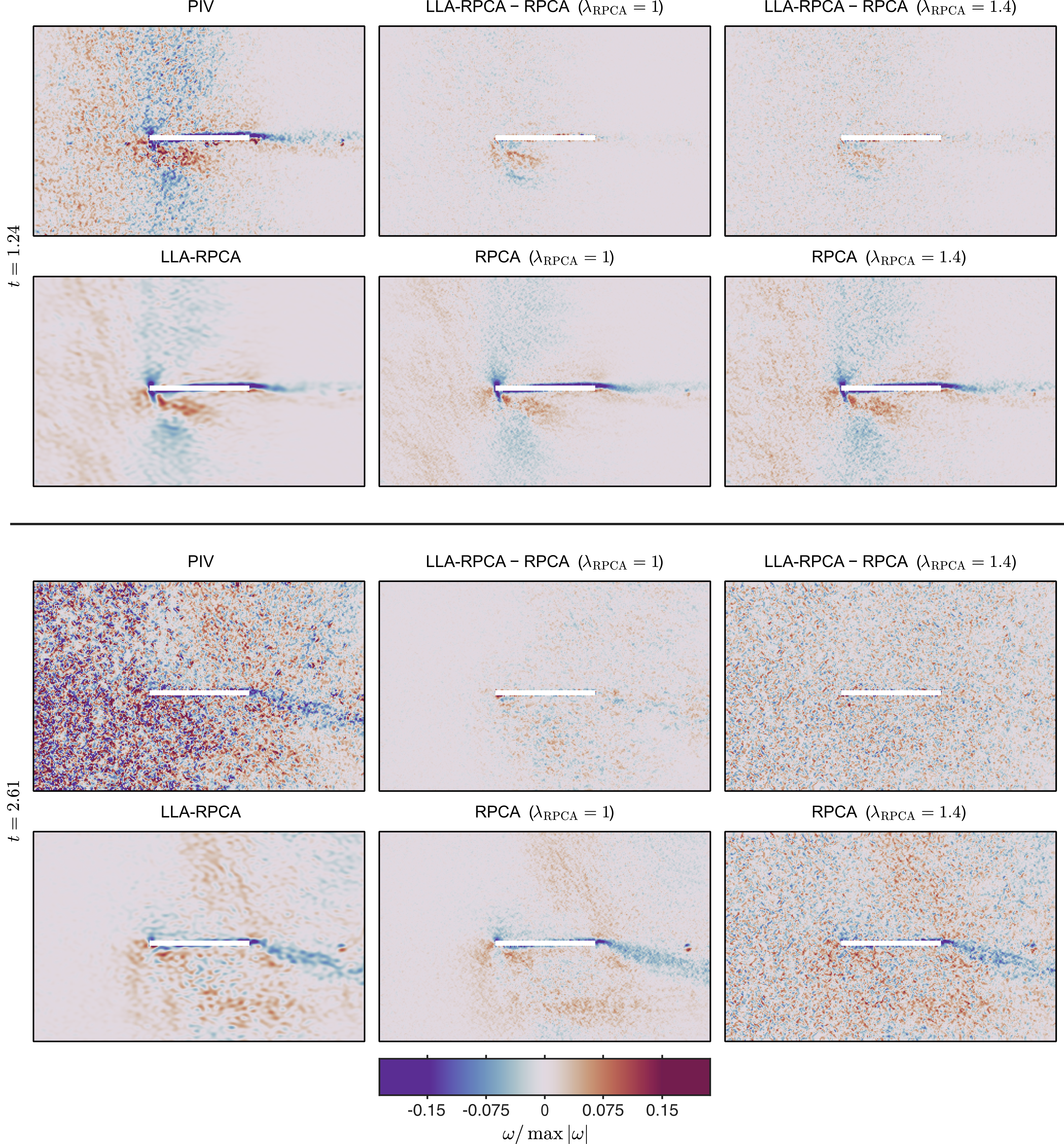}
    \caption{
    Spanwise vorticity fields computed for the experimentally measured transverse extreme gust encounter at \(t=1.24\) and \(t=2.61\).
    For each snapshot, shown are the vorticity computed directly from the measured PIV velocity fields, the vorticity computed from the LLA-RPCA-denoised velocity fields, the vorticity computed from the RPCA-denoised velocity fields with \(\lambda_{\mathrm{RPCA}}=1\) and \(1.4\), and the corresponding differences between the LLA-RPCA- and RPCA-derived vorticity fields.
    }
    \label{fig:Plate_omega}
\end{figure}

RPCA and LLA-RPCA are applied to the measured streamwise and transverse velocity components, \(u\) and \(v\), after which the reconstructed components are used to compute the spanwise vorticity.
The reconstructions are examined at \(t=1.24\) and \(t=2.61\), representing snapshots with substantially different levels of visible PIV artifacts.
For RPCA, both snapshots are reconstructed using \(\lambda_{\mathrm{RPCA}}=1\) and \(1.4\) to determine whether a single tuning factor can simultaneously suppress the PIV artifacts and preserve the underlying flow structures.
For LLA-RPCA, a single reconstruction is used with \(r=16\) and a raw candidate library containing \(6000\) trigonometric and \(100\) graph-Laplacian candidates.
For brevity, only the streamwise velocity component \(u\) is shown in figure~\ref{fig:Plate_u}, which contains the reconstructions at both \(t=1.24\) and \(t=2.61\).
The transverse component \(v\) is denoised in the same manner, using the same parameter settings as for \(u\), and both reconstructed velocity components are then used to compute the spanwise vorticity fields shown for both snapshots in figure~\ref{fig:Plate_omega}.

The streamwise-velocity reconstructions at \(t=1.24\), where the measured PIV field does not exhibit pronounced PIV artifacts, are shown in the upper set of panels in figure~\ref{fig:Plate_u}.
For \(\lambda_{\mathrm{RPCA}}=1\), the RPCA reconstruction reduces the magnitude of the coherent velocity structures below the leading edge of the plate.
In particular, attenuation is visible both in the negative-velocity region directly below the leading edge and in the adjacent positive-velocity region below the plate.
The LLA-RPCA reconstruction remains closer to the measured PIV field in these regions, and the corresponding difference field highlights the loss of magnitude in the RPCA solution.
The absence of artifacts in these regions, together with the coherent spatial structure, indicates that the reduction produced by RPCA is consistent with attenuation of physical flow content rather than removal of PIV artifacts.
This behavior agrees with the attenuation observed for the previous validation datasets.
Increasing the RPCA tuning factor to \(\lambda_{\mathrm{RPCA}}=1.4\) reduces this attenuation.
The coherent velocity structures are retained more closely, and the difference from LLA-RPCA decreases in the regions below the leading edge where the strongest attenuation occurs for \(\lambda_{\mathrm{RPCA}}=1\).
Some discrepancy remains, indicating that attenuation has not been eliminated completely, but the higher value of \(\lambda_{\mathrm{RPCA}}\) improves preservation of the coherent velocity field.

The lower set of panels in figure~\ref{fig:Plate_u} shows the substantially more contaminated snapshot at \(t=2.61\).
The measured PIV field contains widespread PIV artifacts, predominantly in the left part of the domain.
RPCA with \(\lambda_{\mathrm{RPCA}}=1\) removes these artifacts effectively, and its reconstruction agrees closely with the LLA-RPCA reconstruction over most of the domain, as reflected by the comparatively small difference between them.
When the RPCA tuning factor is increased to \(\lambda_{\mathrm{RPCA}}=1.4\), substantial PIV artifacts instead remain in the RPCA reconstruction.
The corresponding difference from the unchanged LLA-RPCA reconstruction therefore becomes much larger throughout the affected regions.
The two velocity snapshots thus reveal the parameter trade-off of standard RPCA.
With \(\lambda_{\mathrm{RPCA}}=1\), RPCA suppresses the strong PIV artifacts at \(t=2.61\) but attenuates coherent flow structures at \(t=1.24\).
Increasing \(\lambda_{\mathrm{RPCA}}\) to \(1.4\) reduces this attenuation, but does not eliminate it, while substantial PIV artifacts already remain in the more contaminated snapshot.
Eliminating the remaining attenuation would require a further increase in \(\lambda_{\mathrm{RPCA}}\), which would make assignment of content to the sparse component still more expensive and thereby further weaken the removal of the PIV artifacts.
Across the displayed snapshots, standard RPCA therefore cannot simultaneously eliminate attenuation of the coherent flow at \(t=1.24\) and suppress the strong PIV artifacts at \(t=2.61\).
LLA-RPCA does not exhibit this trade-off in the displayed velocity fields, retaining the coherent structures at \(t=1.24\) while also suppressing the strong PIV artifacts at \(t=2.61\).

We then examine how these differences in the reconstructed velocity components propagate into the derived spanwise vorticity.
For the PIV result, the vorticity is computed directly from the original measured \(u\) and \(v\) fields without prior denoising.
The RPCA- and LLA-RPCA-derived vorticity fields are instead computed from the corresponding denoised \(u\) and \(v\) components.

The upper set of panels in figure~\ref{fig:Plate_omega} shows how the velocity-field reconstruction behavior at \(t=1.24\) propagates into the spanwise vorticity.
For \(\lambda_{\mathrm{RPCA}}=1\), the attenuation observed in the RPCA velocity reconstruction produces a corresponding reduction in the magnitude of the vorticity structures below the leading edge.
The difference between the RPCA- and LLA-RPCA-derived fields is concentrated in the same region in which their velocity reconstructions differ, showing that the velocity-field discrepancy directly affects the subsequently calculated vorticity.
Increasing \(\lambda_{\mathrm{RPCA}}\) to \(1.4\) reduces this discrepancy.
The RPCA-derived vorticity agrees more closely with the LLA-RPCA-derived field, and the difference between them is correspondingly smaller in the coherent structures below the leading edge.
The vorticity comparison therefore reflects the same reduction in attenuation observed in the streamwise-velocity reconstruction when the RPCA tuning factor is increased.

The lower set of panels in figure~\ref{fig:Plate_omega} presents the corresponding comparison for the strongly contaminated snapshot at \(t=2.61\).
Computing vorticity directly from the measured PIV velocities strongly amplifies the localized velocity errors, producing widespread high-magnitude artifacts throughout the PIV-derived vorticity field.
RPCA with $\lambda_{\mathrm{RPCA}}=1$ suppresses these artifacts in the velocity fields before differentiation, and the resulting vorticity field agrees closely with the LLA-RPCA-derived field, with only minor differences between them.
For $\lambda_{\mathrm{RPCA}}=1.4$, however, the PIV artifacts retained in the RPCA velocity fields are strongly amplified by spatial differentiation, producing pronounced artifacts throughout the resulting vorticity field.
The LLA-RPCA-derived vorticity remains free of these artifacts, and the corresponding difference field clearly identifies the artifacts that have propagated from the RPCA velocity reconstruction into the derived quantity.
This sensitivity to the preceding velocity reconstruction is consistent with the NACA0012 results, where errors were also shown to propagate through the derivative operation into the calculated vorticity fields.

Together, figures~\ref{fig:Plate_u} and \ref{fig:Plate_omega} therefore show the same RPCA tuning-factor trade-off at velocity and derived-vorticity levels.
The lower value $\lambda_{\mathrm{RPCA}}=1$ is required to suppress the strong PIV artifacts in the more contaminated snapshot, but produces attenuation of coherent flow content in the less contaminated snapshot.
Increasing $\lambda_{\mathrm{RPCA}}$ to $1.4$ reduces this attenuation but allows substantial PIV artifacts to remain, which are subsequently amplified in the calculated vorticity.
This behavior reflects the opposing RPCA failure modes described at the beginning of section~\ref{sec:method}.
A lower \(\lambda_{\mathrm{RPCA}}\) makes transfer of physical content into the sparse component less expensive and therefore promotes attenuation, whereas increasing \(\lambda_{\mathrm{RPCA}}\) makes assignment of PIV artifacts to the sparse component more expensive and consequently promotes their retention in the recovered low-rank field.
LLA-RPCA does not exhibit this trade-off in the displayed snapshots considered here.
It retains the coherent velocity and derived-vorticity structures at $t=1.24$ while also suppressing the strong PIV artifacts at $t=2.61$.
Although the absence of an uncorrupted experimental reference prevents direct reconstruction-error evaluation, the velocity and vorticity comparisons indicate that LLA-RPCA more consistently separates coherent flow content from the naturally occurring PIV artifacts in this experimental case.

In addition to the spatial comparisons above, we examine the probability-density distributions of the field values before and after applying the denoising methods.
For each of the two displayed snapshots, the distributions are calculated over the valid spatial domain for the measured PIV field, the LLA-RPCA reconstruction, and the two RPCA reconstructions.
The velocity distributions provide a measure of how the denoising changes the range and relative occurrence of the reconstructed streamwise-velocity values, and can indicate attenuation when the reconstructed distribution contracts toward the center relative to the measured field.
The vorticity distributions provide a corresponding view of the derived quantity and are sensitive to the high-magnitude values produced when localized velocity-field errors are amplified by spatial differentiation.
These distributions do not retain spatial information, but provide a complementary statistical comparison of the reconstruction behavior identified in figures~\ref{fig:Plate_u} and \ref{fig:Plate_omega}.

\begin{figure}[t]
\vspace{-2mm}
    \centering
    \includegraphics[width=0.95\linewidth]{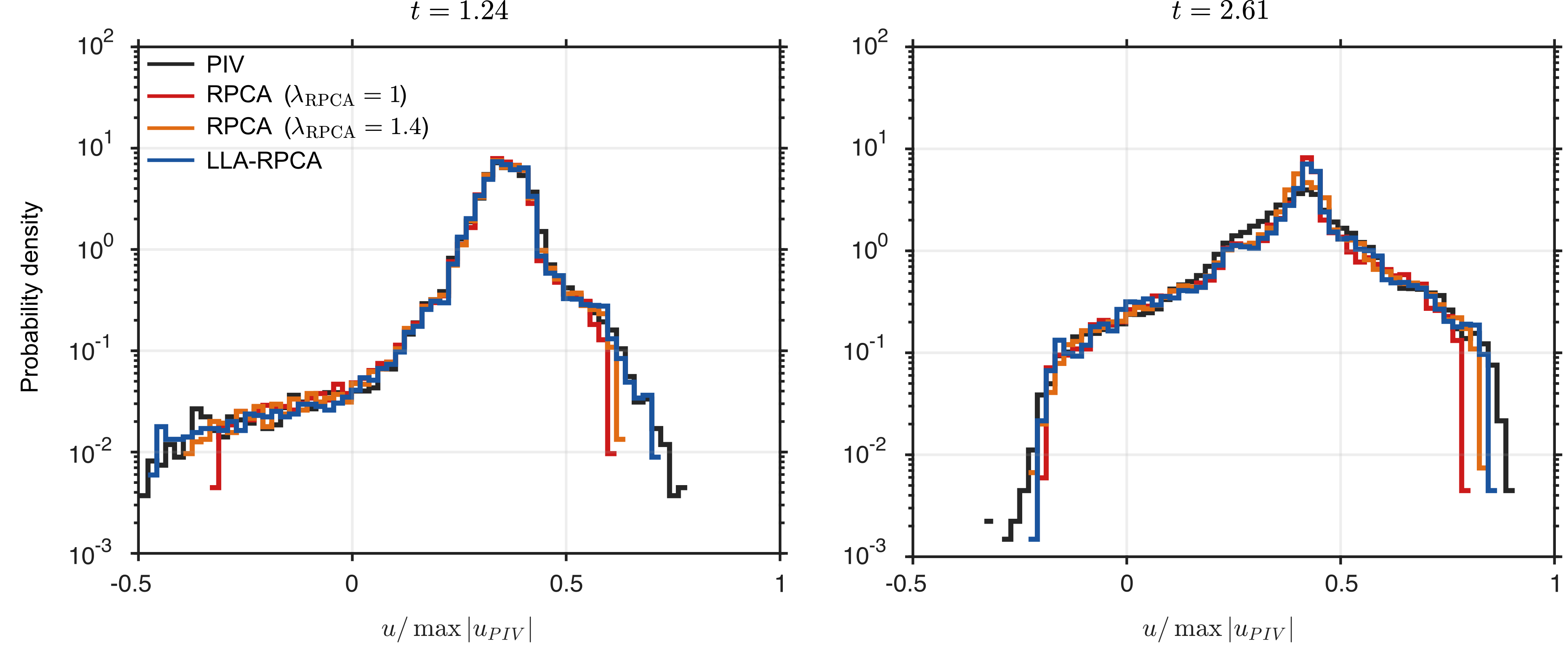}
    \caption{
    Probability-density distributions of the normalized streamwise velocity for the experimentally measured transverse extreme gust encounter at \(t=1.24\) and \(t=2.61\).
    Shown are the distributions obtained from the measured PIV field, the LLA-RPCA reconstruction, and the RPCA reconstructions with \(\lambda_{\mathrm{RPCA}}=1\) and \(1.4\).
    }
    \label{fig:Plate_u_pdf}
\end{figure}

The streamwise-velocity distributions in figure~\ref{fig:Plate_u_pdf} show how the attenuation and artifact-removal behavior identified from the spatial fields is reflected in the distribution of velocity values.
At \(t=1.24\), the measured PIV field contains no pronounced visible artifacts, so the distribution provides a useful reference for assessing the attenuation observed in the corresponding panels of figure~\ref{fig:Plate_u}.
LLA-RPCA follows the measured PIV distribution most closely in the positive and negative tails.
The RPCA distribution with \(\lambda_{\mathrm{RPCA}}=1.4\) contracts toward the center and terminates at smaller velocity magnitudes, while the distribution for \(\lambda_{\mathrm{RPCA}}=1\) contracts still further.
The ordering of the distribution widths therefore agrees with the spatial reconstruction: RPCA with \(\lambda_{\mathrm{RPCA}}=1\) exhibits the strongest attenuation, increasing the tuning factor to \(1.4\) reduces this attenuation, and LLA-RPCA preserves the measured velocity range most closely.

At \(t=2.61\), the reconstructed velocity distributions retain the same ordering in their tails, with LLA-RPCA extending to the largest velocity magnitudes, followed by RPCA with \(\lambda_{\mathrm{RPCA}}=1.4\) and then RPCA with \(\lambda_{\mathrm{RPCA}}=1\).
A second feature appears near the dominant peak at \(u/\max|u_{\mathrm{PIV}}|\approx0.42\), corresponding to the approximately uniform streamwise velocity in the far field.
At \(t=1.24\), where this region is comparatively free of visible artifacts, the measured PIV field produces a pronounced peak at this value.
At \(t=2.61\), widespread corruption in the left part of the measured field reduces the number of entries retaining this characteristic velocity, and the corresponding peak in the measured PIV distribution is substantially lower.
LLA-RPCA and RPCA with \(\lambda_{\mathrm{RPCA}}=1\) recover a higher peak whose magnitude remains close to that observed at \(t=1.24\), consistent with the effective suppression of the widespread artifacts seen at \(t=2.61\) in figure~\ref{fig:Plate_u}.
The RPCA distribution for \(\lambda_{\mathrm{RPCA}}=1.4\), in contrast, remains closer to the measured PIV distribution and exhibits a lower peak.
This is consistent with the substantial PIV artifacts retained in the corresponding RPCA reconstruction.

\begin{figure}[t]
\vspace{-2mm}
    \centering
    \includegraphics[width=0.95\linewidth]{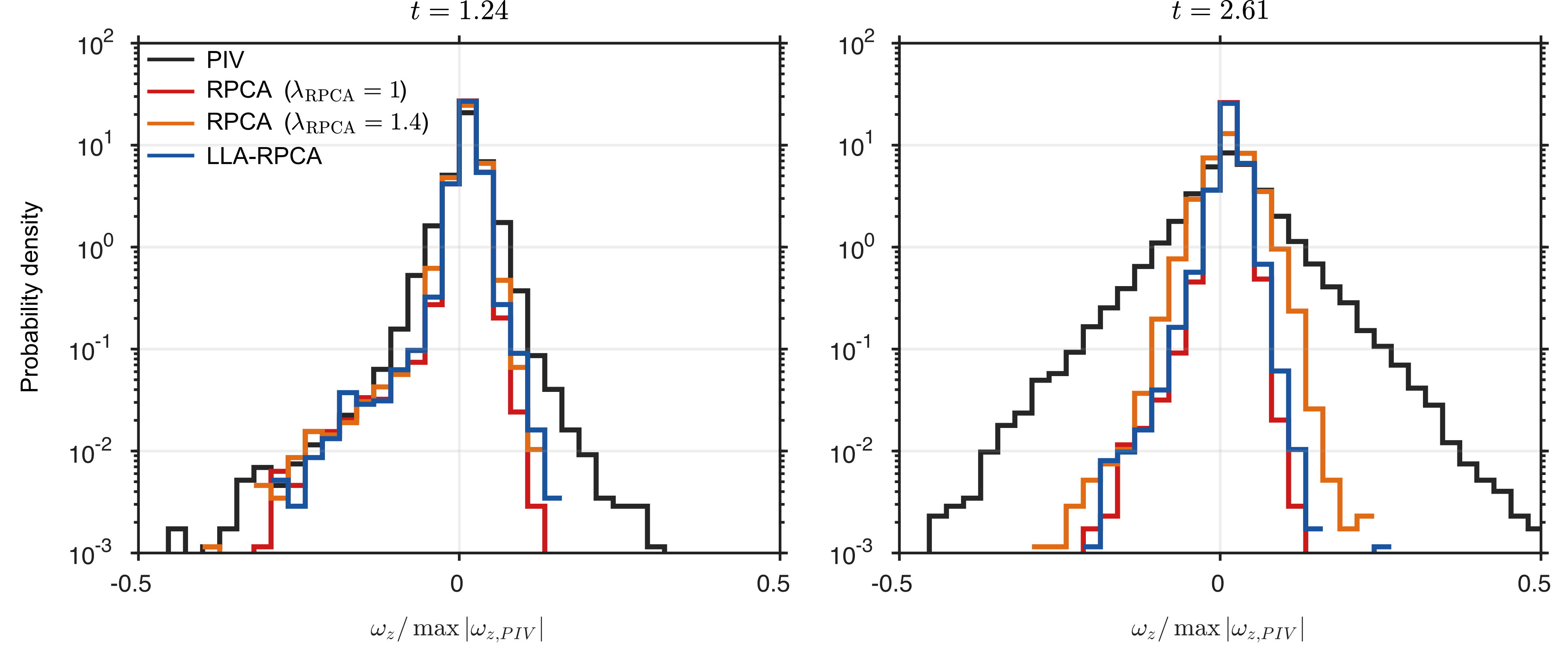}
    \caption{
    Probability-density distributions of the normalized spanwise vorticity for the experimentally measured transverse extreme gust encounter at \(t=1.24\) and \(t=2.61\).
    The measured-PIV distribution is obtained from vorticity computed directly from the original measured velocity components, whereas the RPCA and LLA-RPCA distributions are obtained from vorticity computed from the corresponding denoised velocity components.
    }
    \label{fig:Plate_omega_pdf}
\end{figure}

The corresponding spanwise-vorticity distributions in figure~\ref{fig:Plate_omega_pdf} show how the velocity-field reconstruction behavior propagates into the derived quantity.
At \(t=1.24\), the negative-vorticity tail does not exhibit a clear ordering among the reconstructed distributions.
On the positive side, however, the LLA-RPCA distribution remains closest to the measured-PIV distribution, while RPCA with \(\lambda_{\mathrm{RPCA}}=1.4\) is shifted farther toward zero and RPCA with \(\lambda_{\mathrm{RPCA}}=1\) contracts still more strongly.
This ordering is consistent with the attenuation identified in the velocity fields and in the \(t=1.24\) panels of figure~\ref{fig:Plate_omega}.
The stronger velocity attenuation produced by RPCA with \(\lambda_{\mathrm{RPCA}}=1\) results in the largest reduction in derived vorticity magnitude, while increasing \(\lambda_{\mathrm{RPCA}}\) to \(1.4\) reduces this effect.
The LLA-RPCA distribution also lies somewhat inside the measured-PIV distribution.
This is consistent with the fact that the measured-PIV vorticity is calculated directly from the raw velocity components, so small-scale measurement variations that remain in those fields can contribute additional high-magnitude vorticity values after spatial differentiation.

At \(t=2.61\), the measured-PIV vorticity distribution has substantially broader positive and negative tails.
This agrees with the \(t=2.61\) panels of figure~\ref{fig:Plate_omega}, where the widespread velocity-field errors are strongly amplified by spatial differentiation and produce high-magnitude vorticity artifacts.
All three denoised reconstructions produce considerably narrower distributions, showing that each removes part of this artifact-dominated high-magnitude content.
The LLA-RPCA and RPCA distributions for \(\lambda_{\mathrm{RPCA}}=1\) are similar and substantially narrower than the measured-PIV distribution, consistent with the effective artifact suppression observed in the corresponding reconstructed fields.
The RPCA distribution for \(\lambda_{\mathrm{RPCA}}=1.4\), however, remains visibly broader and more similar to the measured-PIV distribution.
This is consistent with the PIV artifacts retained in the RPCA velocity reconstruction with \(\lambda_{\mathrm{RPCA}}=1.4\), which are subsequently amplified in the derived vorticity field.

The probability-density comparisons therefore provide complementary statistical support for the spatial-field observations.
At the less contaminated snapshot, the distributions reflect the progressive reduction in attenuation as \(\lambda_{\mathrm{RPCA}}\) is increased from \(1\) to \(1.4\), with LLA-RPCA preserving the measured velocity distribution most closely.
At the more contaminated snapshot, they show the opposite consequence of the same RPCA tuning-factor change: the larger value retains more of the artifact-contaminated PIV distribution, whereas LLA-RPCA and RPCA with \(\lambda_{\mathrm{RPCA}}=1\) more effectively recover the dominant velocity distribution and suppress the broad vorticity tails associated with the PIV artifacts.
The probability-density results therefore reinforce the tuning-factor trade-off identified from the velocity and vorticity fields.

\section{Concluding remarks}
\label{sec:conc}

LLA-RPCA was developed to recover coherent flow fields and dominant POD subspaces under severe large-amplitude entrywise corruption.
Unlike standard RPCA, which determines the recovered low-rank representation through a balance between nuclear-norm and entrywise \(\ell_1\)-norm penalties, LLA-RPCA introduces two main modifications.
The candidate-library restriction limits the admissible spatial patterns and therefore reduces the ability of the recovered component to represent irregular corruption when this corruption is less compatible with the library span than the coherent flow.
The explicit low-rank factorization with prescribed modal capacity replaces nuclear-norm control of the recovered component, removing the direct nuclear-norm incentive for physical-content attenuation while controlling the number of spatial directions available to represent the flow.

For the NACA0012 wake, LLA-RPCA suppressed the imposed large-amplitude entrywise corruption while preserving the wake structure, vorticity magnitude, and dominant POD subspace over most of the tested corruption range.
The total-field and fluctuation errors remained nearly constant through \(\eta=0.7\), and the total-field error remained lower than the RPCA value through \(\eta=0.8\).
At the largest corruption fractions, the limited number of informative entries caused the LLA-RPCA errors to increase sharply.
RPCA instead showed progressively increasing reconstruction errors, attenuation of the reconstructed vorticity, residual noise in the higher-order POD modes, and deterioration of the recovered POD subspace as the corruption fraction increased.
LLA-RPCA therefore retained a coherent reconstruction over a substantially wider corruption range than standard RPCA for this case.

For the oscillating-cylinder case, LLA-RPCA produced lower total-field and fluctuation errors than RPCA at every tested corruption fraction.
Both errors increased only slightly as the corruption fraction rose, consistent with the limited degradation observed in the reconstructed wake sequence.
LLA-RPCA retained the instantaneous vortex positions, wake envelope, and multidimensional POD subspace under severe imposed corruption.
RPCA instead progressively lost the vortex-arm structure as the corruption fraction increased.
At lower corruption fractions, this degradation included attenuation and loss of instantaneous wake structure, whereas at \(\eta=0.9\) the reconstruction collapsed to an almost spatially and temporally uniform field at the imposed positive corruption level of \(10\sigma\), demonstrating substantial contamination of the recovered low-rank component.
For the displayed case at \(\eta=0.9\), the RPCA low-rank matrix had rank one, so only its first POD direction and first subspace-alignment value were defined.
The higher-order modal directions required to represent the changing wake were absent.
At low corruption fractions, RPCA produced a lower SSIM error because it retained thin seeding-related streaklines and fine background structure that were smoothed by the finite LLA-RPCA library.
This lower SSIM error reflected more accurate reproduction of small-scale image detail rather than more accurate recovery of the instantaneous wake evolution.
As the corruption fraction increased and the RPCA wake structure deteriorated, LLA-RPCA also obtained a lower SSIM error.

The experimentally measured transverse extreme gust case exposed a different limitation of standard RPCA.
For the lower tuning factor \(\lambda_{\mathrm{RPCA}}=1\), RPCA suppressed the strong PIV artifacts in the more contaminated snapshot but attenuated coherent velocity structures in a snapshot without pronounced visible artifacts.
This attenuation propagated into the spanwise vorticity subsequently calculated from the reconstructed velocity components.
Increasing the tuning factor to \(\lambda_{\mathrm{RPCA}}=1.4\) reduced the attenuation, but did not eliminate it, while substantial PIV artifacts already remained in the RPCA reconstruction and were amplified by spatial differentiation.
The probability-density distributions support the same interpretation.
At \(t=1.24\), the RPCA velocity distributions contract progressively toward the center as \(\lambda_{\mathrm{RPCA}}\) is reduced from \(1.4\) to \(1\), consistent with increasing attenuation, while LLA-RPCA remains closest to the measured velocity distribution.
At \(t=2.61\), LLA-RPCA and RPCA with \(\lambda_{\mathrm{RPCA}}=1\) recover the dominant streamwise-velocity peak more strongly and produce substantially narrower vorticity distributions than the measured PIV field, whereas RPCA with \(\lambda_{\mathrm{RPCA}}=1.4\) remains closer to the artifact-contaminated PIV distributions.
A further increase in \(\lambda_{\mathrm{RPCA}}\) would be required to remove the remaining attenuation, while simultaneously making assignment of the PIV artifacts to the sparse component still more expensive.
Across the displayed snapshots, the results therefore show that standard RPCA cannot simultaneously eliminate attenuation of the coherent flow at \(t=1.24\) and suppress the strong PIV artifacts at \(t=2.61\).
LLA-RPCA does not exhibit this trade-off in the displayed snapshots.
It preserves the coherent velocity and derived-vorticity structures in the less contaminated snapshot while also suppressing the strong PIV artifacts in the more contaminated snapshot.
Because no uncorrupted experimental reference is available, these findings remain qualitative, but the combined spatial-field and probability-density comparisons indicate that LLA-RPCA more consistently separates coherent flow content from the naturally occurring PIV artifacts.

The performance of LLA-RPCA depends on the prescribed candidate library and modal capacity.
Because the recovered low-rank field is restricted to the span of the candidate library, coherent structures outside this span cannot be represented accurately and may be smoothed or assigned partly to the corruption component.
This limitation was visible in the oscillating-cylinder case, where the finite library did not resolve the thin seeding-related streaklines and consequently produced a smoother reconstruction than the reference data.
The prescribed modal capacity is also case-dependent.
Too little modal capacity restricts the coherent flow configurations that can be represented, whereas excessive modal capacity increases the flexibility of the low-rank component and may allow it to represent more corruption.
The selected modal capacity must therefore balance flow representation against corruption exclusion.

The bilinear factorization of the low-rank component makes the optimization jointly nonconvex in \(\mathbf{A}\) and \(\mathbf{B}\).
Convergence does not establish global optimality, and the recovered decomposition may depend on the candidate library, prescribed modal capacity, initial learned subspace, and continuation schedule.

The formulation imposes no temporal regularity on the coefficient matrix \(\mathbf{A}\).
This permits transient and nonperiodic evolution to be represented without prescribing a temporal model, but it also allows individual coefficient vectors to become poorly constrained when too few informative observations remain in a snapshot.

The controlled validation considered replacement corruption with one fixed amplitude of \(10\sigma\).
The NACA0012 case included both randomly distributed and shear-based mixed-sign corruption, while the cylinder case used randomly distributed positive corruption.
The results therefore do not establish performance under temporally persistent, correlated, lower-amplitude, colored, biased, or missing-data corruption, nor do they separately evaluate dense low-amplitude noise.
The study also did not compare LLA-RPCA with other prescribed-rank or structured robust-decomposition methods or evaluate its influence on downstream analyses beyond POD and the calculation of derived vorticity fields.
The absence of an uncorrupted reference for the flat-plate case further limits the extent to which physical structures and measurement artifacts can be distinguished quantitatively.

Future work should first establish systematic procedures for selecting the candidate library and prescribed modal capacity.
Ablation studies of trigonometric, graph-Laplacian, combined, and alternative analytical or data-driven libraries could quantify how candidate-family composition and library size affect reconstruction and POD subspace recovery.
Modal-capacity selection should balance coherent-flow representation against the additional flexibility available to represent corruption.

Temporal regularization could be examined as a means of stabilizing individual coefficient vectors when few reliable observations remain, while preserving rapid transients.
The sensitivity of the nonconvex optimization also requires evaluation across multiple initial learned subspaces and continuation schedules.

Broader controlled validation should vary corruption amplitude, spatial arrangement, temporal occurrence, sign balance, bias, correlation, dense low-amplitude noise, and missing data.
Comparisons with other prescribed-rank or structured robust-decomposition methods would clarify the relative benefits of the proposed formulation.
The effects of LLA-RPCA preprocessing on DMD and reduced-order modeling also remain to be assessed.

Across the two controlled corruption cases and the experimental gust encounter, LLA-RPCA more consistently suppresses corrupted content and naturally occurring PIV artifacts while preserving coherent flow structures and the modal directions required to represent the underlying dynamics when the candidate library and prescribed modal capacity represent the relevant spatial content adequately.

\section*{Acknowledgments}

K.F. acknowledges support from the JSPS KAKENHI Grant No.~JP25K23418 and No.~JP26K01129, the JST PRESTO Grant No.~JPMJPR25KA, and the MEXT Coordination Funds for Promoting Aerospace Utilization Grant No.~JPJ000959.

\section*{Data availability}

An example implementation of LLA-RPCA for the numerically simulated NACA0012 case is publicly available online~(\url{https://github.com/kfukami/Library-RPCA}).
The oscillating-cylinder video data used in this study are publicly available in a Zenodo repository~\cite{tomasetto2024dataset}.
The transverse-gust PIV data are publicly available in the University of Michigan Deep Blue Data repository~\cite{towne2022gustdataset}.

\appendix

\section{Derivation of the reduced augmented Lagrangian}
\label{app:reduced_augmented_lagrangian}

This section derives the reduced augmented Lagrangian used for the alternating update equations.
We define the decomposition residual as
\begin{equation}
\mathbf{R}
=
\mathbf{X}
-
\boldsymbol{\Psi}\mathbf{B}\mathbf{A}
-
\mathbf{S}.
\label{eq:app_residual_definition}
\end{equation}
Using this definition, the two constraint-enforcement terms in equation~\eqref{eq:augmented_lagrangian_initial} are
\begin{equation}
\left\langle
\mathbf{Y},
\mathbf{R}
\right\rangle
+
\frac{\nu}{2}
\left\|
\mathbf{R}
\right\|_F^2.
\label{eq:app_constraint_terms}
\end{equation}
These terms can be combined by completing the square.
We first consider
\begin{equation}
\frac{\nu}{2}
\left\|
\mathbf{R}
+
\frac{1}{\nu}\mathbf{Y}
\right\|_F^2.
\label{eq:app_square_term}
\end{equation}
Using the Frobenius-norm identity
\begin{equation}
\left\|
\mathbf{U}
+
\mathbf{V}
\right\|_F^2
=
\left\|
\mathbf{U}
\right\|_F^2
+
2
\left\langle
\mathbf{U},
\mathbf{V}
\right\rangle
+
\left\|
\mathbf{V}
\right\|_F^2,
\label{eq:app_frobenius_identity}
\end{equation}
equation~\eqref{eq:app_square_term} expands as
\begin{align}
\frac{\nu}{2}
\left\|
\mathbf{R}
+
\frac{1}{\nu}\mathbf{Y}
\right\|_F^2
&=
\frac{\nu}{2}
\left(
\left\|
\mathbf{R}
\right\|_F^2
+
2
\left\langle
\mathbf{R},
\frac{1}{\nu}\mathbf{Y}
\right\rangle
+
\left\|
\frac{1}{\nu}\mathbf{Y}
\right\|_F^2
\right)
\nonumber\\
&=
\frac{\nu}{2}
\left\|
\mathbf{R}
\right\|_F^2
+
\left\langle
\mathbf{R},
\mathbf{Y}
\right\rangle
+
\frac{1}{2\nu}
\left\|
\mathbf{Y}
\right\|_F^2.
\label{eq:app_square_expansion}
\end{align}
Since the Frobenius inner product is symmetric, $\left\langle\mathbf{R},\mathbf{Y}\right\rangle=\left\langle\mathbf{Y},\mathbf{R}\right\rangle$.
Rearranging equation~\eqref{eq:app_square_expansion} gives
\begin{equation}
\left\langle
\mathbf{Y},
\mathbf{R}
\right\rangle
+
\frac{\nu}{2}
\left\|
\mathbf{R}
\right\|_F^2
=
\frac{\nu}{2}
\left\|
\mathbf{R}
+
\frac{1}{\nu}\mathbf{Y}
\right\|_F^2
-
\frac{1}{2\nu}
\left\|
\mathbf{Y}
\right\|_F^2.
\label{eq:app_square_completion_identity}
\end{equation}
Substituting equation~\eqref{eq:app_square_completion_identity} into equation~\eqref{eq:augmented_lagrangian_initial} yields
\begin{align}
\mathcal{L}_{\nu}
\left(
\mathbf{A},
\mathbf{B},
\mathbf{S},
\mathbf{Y}
\right)
=
&\;
\left\|
\mathbf{S}
\right\|_1
+
\frac{\nu}{2}
\left\|
\mathbf{X}
-
\boldsymbol{\Psi}\mathbf{B}\mathbf{A}
-
\mathbf{S}
+
\frac{1}{\nu}\mathbf{Y}
\right\|_F^2
\nonumber\\
&
-
\frac{1}{2\nu}
\left\|
\mathbf{Y}
\right\|_F^2.
\label{eq:app_augmented_lagrangian_completed_square}
\end{align}
When deriving the updates for $\mathbf{A}$, $\mathbf{B}$, and $\mathbf{S}$, the quantities $\mathbf{Y}$ and $\nu$ are held fixed.
Therefore, the final term in equation~\eqref{eq:app_augmented_lagrangian_completed_square} is independent of $\mathbf{A}$, $\mathbf{B}$, and $\mathbf{S}$.
It can therefore be discarded without changing the minimizer with respect to these variables.
This gives the reduced augmented Lagrangian
\begin{equation}
\mathcal{L}_{\nu}^{\mathrm{red}}
\left(
\mathbf{A},
\mathbf{B},
\mathbf{S},
\mathbf{Y}
\right)
=
\left\|
\mathbf{S}
\right\|_1
+
\frac{\nu}{2}
\left\|
\mathbf{X}
-
\boldsymbol{\Psi}\mathbf{B}\mathbf{A}
-
\mathbf{S}
+
\frac{1}{\nu}\mathbf{Y}
\right\|_F^2.
\label{eq:app_augmented_lagrangian_reduced}
\end{equation}
This is the form used to derive the alternating update equations for $\mathbf{A}$, $\mathbf{B}$, and $\mathbf{S}$.

\section{Derivation of the alternating update equations}
\label{app:alternating_updates}

The alternating update equations are obtained from the reduced augmented Lagrangian in equation~\eqref{eq:augmented_lagrangian_reduced}.
During each update, the remaining variables and the current Lagrange multiplier matrix are held fixed.

\subsection{Update for snapshot-dependent modal coefficients \texorpdfstring{$\mathbf{A}$}{A}}
\label{app:update_A}

For the $\mathbf{A}$-update, $\mathbf{B}$, $\mathbf{S}$, and $\mathbf{Y}$ are fixed.
Let
\begin{equation}
\boldsymbol{\Phi}^{(h)}
=
\boldsymbol{\Psi}\mathbf{B}^{(h)},
\qquad
\mathbf{Z}_A^{(h)}
=
\mathbf{X}
-
\mathbf{S}^{(h)}
+
\frac{1}{\nu^{(h)}}
\mathbf{Y}^{(h)}.
\end{equation}
The $\mathbf{A}$-subproblem is then
\begin{equation}
\mathbf{A}^{(h+1)}
=
\arg\min_{\mathbf{A}}
\frac{\nu^{(h)}}{2}
\left\|
\mathbf{Z}_A^{(h)}
-
\boldsymbol{\Phi}^{(h)}
\mathbf{A}
\right\|_F^2.
\label{eq:app_A_subproblem}
\end{equation}
Taking the derivative with respect to $\mathbf{A}$ gives
\begin{equation}
\nabla_{\mathbf{A}}J_A
=
-
\nu^{(h)}
\left(
\boldsymbol{\Phi}^{(h)}
\right)^T
\mathbf{Z}_A^{(h)}
+
\nu^{(h)}
\left(
\boldsymbol{\Phi}^{(h)}
\right)^T
\boldsymbol{\Phi}^{(h)}
\mathbf{A}.
\end{equation}
Setting this gradient equal to zero yields
\begin{equation}
\left(
\boldsymbol{\Phi}^{(h)}
\right)^T
\boldsymbol{\Phi}^{(h)}
\mathbf{A}^{(h+1)}
=
\left(
\boldsymbol{\Phi}^{(h)}
\right)^T
\mathbf{Z}_A^{(h)}.
\end{equation}
Because $\boldsymbol{\Phi}^{(h)}$ has orthonormal columns,
\begin{equation}
\left(
\boldsymbol{\Phi}^{(h)}
\right)^T
\boldsymbol{\Phi}^{(h)}
=
\mathbf{I}_r.
\end{equation}
Substituting the definition of $\mathbf{Z}_A^{(h)}$ gives
\begin{equation}
\mathbf{A}^{(h+1)}
=
\left(
\boldsymbol{\Phi}^{(h)}
\right)^T
\left(
\mathbf{X}
-
\mathbf{S}^{(h)}
+
\frac{1}{\nu^{(h)}}
\mathbf{Y}^{(h)}
\right).
\label{eq:app_A_update_final}
\end{equation}

\subsection{Update for mode-combination coefficients \texorpdfstring{$\mathbf{B}$}{B}}
\label{app:update_B}

For the $\mathbf{B}$-update, $\mathbf{A}$, $\mathbf{S}$, and $\mathbf{Y}$ are fixed.
Defining
\begin{equation}
\mathbf{Z}_B^{(h)}
=
\mathbf{X}
-
\mathbf{S}^{(h)}
+
\frac{1}{\nu^{(h)}}
\mathbf{Y}^{(h)},
\label{eq:app_Z_B_definition}
\end{equation}
the $\mathbf{B}$-subproblem is
\begin{equation}
\mathbf{B}^{(h+1)}
=
\arg\min_{\mathbf{B}}
\frac{\nu^{(h)}}{2}
\left\|
\mathbf{Z}_B^{(h)}
-
\boldsymbol{\Psi}
\mathbf{B}
\mathbf{A}^{(h+1)}
\right\|_F^2.
\label{eq:app_B_subproblem}
\end{equation}
Taking the derivative with respect to $\mathbf{B}$ gives
\begin{align}
\nabla_{\mathbf{B}}J_B
={}&
\nu^{(h)}
\boldsymbol{\Psi}^{T}
\left(
\boldsymbol{\Psi}
\mathbf{B}
\mathbf{A}^{(h+1)}
-
\mathbf{Z}_B^{(h)}
\right)
\left(
\mathbf{A}^{(h+1)}
\right)^T
\nonumber\\
={}&
-
\nu^{(h)}
\boldsymbol{\Psi}^{T}
\mathbf{Z}_B^{(h)}
\left(
\mathbf{A}^{(h+1)}
\right)^T
\nonumber\\
&
+
\nu^{(h)}
\boldsymbol{\Psi}^{T}
\boldsymbol{\Psi}
\mathbf{B}
\mathbf{A}^{(h+1)}
\left(
\mathbf{A}^{(h+1)}
\right)^T.
\label{eq:app_B_gradient}
\end{align}
Setting this gradient equal to zero yields
\begin{equation}
\boldsymbol{\Psi}^{T}
\boldsymbol{\Psi}
\mathbf{B}^{(h+1)}
\mathbf{A}^{(h+1)}
\left(
\mathbf{A}^{(h+1)}
\right)^T
=
\boldsymbol{\Psi}^{T}
\mathbf{Z}_B^{(h)}
\left(
\mathbf{A}^{(h+1)}
\right)^T.
\label{eq:app_B_linear_matrix_equation}
\end{equation}
Without orthonormality of the candidate library, equation~\eqref{eq:app_B_linear_matrix_equation} remains coupled through $\boldsymbol{\Psi}^{T}\boldsymbol{\Psi}$.
Because the candidate library is orthonormal,
\begin{equation}
\boldsymbol{\Psi}^{T}
\boldsymbol{\Psi}
=
\mathbf{I}_k,
\label{eq:app_B_Psi_orthonormality}
\end{equation}
this coupling is removed.
Equation~\eqref{eq:app_B_linear_matrix_equation} therefore reduces to
\begin{equation}
\mathbf{B}^{(h+1)}
\mathbf{A}^{(h+1)}
\left(
\mathbf{A}^{(h+1)}
\right)^T
=
\boldsymbol{\Psi}^{T}
\mathbf{Z}_B^{(h)}
\left(
\mathbf{A}^{(h+1)}
\right)^T.
\label{eq:app_B_orthonormal_equation}
\end{equation}
Factoring $\mathbf{B}^{(h+1)}$ from the left-hand side gives
\begin{equation}
\mathbf{B}^{(h+1)}
\left[
\mathbf{A}^{(h+1)}
\left(
\mathbf{A}^{(h+1)}
\right)^T
\right]
=
\boldsymbol{\Psi}^{T}
\mathbf{Z}_B^{(h)}
\left(
\mathbf{A}^{(h+1)}
\right)^T.
\label{eq:app_B_reduced_system}
\end{equation}
The update therefore requires factorizing only the $r\times r$ matrix in equation~\eqref{eq:app_B_reduced_system}, whose dimension is independent of the library size $k$.
Assuming that $\mathbf{A}^{(h+1)}$ has full row rank, the matrix
\begin{equation}
\mathbf{A}^{(h+1)}
\left(
\mathbf{A}^{(h+1)}
\right)^T
\end{equation}
is positive definite, and the solution is
\begin{equation}
\mathbf{B}^{(h+1)}
=
\boldsymbol{\Psi}^{T}
\mathbf{Z}_B^{(h)}
\left(
\mathbf{A}^{(h+1)}
\right)^T
\left[
\mathbf{A}^{(h+1)}
\left(
\mathbf{A}^{(h+1)}
\right)^T
\right]^{-1}.
\label{eq:app_B_solution}
\end{equation}
Substituting the definition of $\mathbf{Z}_B^{(h)}$ gives
\begin{equation}
\mathbf{B}^{(h+1)}
=
\boldsymbol{\Psi}^{T}
\left(
\mathbf{X}
-
\mathbf{S}^{(h)}
+
\frac{1}{\nu^{(h)}}
\mathbf{Y}^{(h)}
\right)
\left(
\mathbf{A}^{(h+1)}
\right)^T
\left[
\mathbf{A}^{(h+1)}
\left(
\mathbf{A}^{(h+1)}
\right)^T
\right]^{-1},
\end{equation}
which is the update reported in section~\ref{subsubsec:update_B}.

\subsection{Update for corruption component \texorpdfstring{$\mathbf{S}$}{S}}
\label{app:update_S}

For the $\mathbf{S}$-update, $\mathbf{A}$, $\mathbf{B}$, and $\mathbf{Y}$ are fixed.
Defining
\begin{equation}
\mathbf{Z}_S^{(h)}
=
\mathbf{X}
-
\boldsymbol{\Psi}
\mathbf{B}^{(h+1)}
\mathbf{A}^{(h+1)}
+
\frac{1}{\nu^{(h)}}
\mathbf{Y}^{(h)},
\end{equation}
the $\mathbf{S}$-subproblem is
\begin{equation}
\mathbf{S}^{(h+1)}
=
\arg\min_{\mathbf{S}}
\left[
\left\|
\mathbf{S}
\right\|_1
+
\frac{\nu^{(h)}}{2}
\left\|
\mathbf{Z}_S^{(h)}
-
\mathbf{S}
\right\|_F^2
\right].
\label{eq:app_S_subproblem}
\end{equation}
This problem separates entrywise.
For each entry, the scalar problem is
\begin{equation}
S_{ij}^{(h+1)}
=
\arg\min_s
\left[
|s|
+
\frac{\nu^{(h)}}{2}
\left(
Z_{S,ij}^{(h)}
-
s
\right)^2
\right].
\end{equation}
Its solution is the soft-thresholding operator
\begin{equation}
S_{ij}^{(h+1)}
=
\operatorname{sign}
\left(
Z_{S,ij}^{(h)}
\right)
\max
\left(
\left|
Z_{S,ij}^{(h)}
\right|
-
\frac{1}{\nu^{(h)}},
0
\right).
\end{equation}
Equivalently, in matrix form,
\begin{equation}
\mathbf{S}^{(h+1)}
=
\mathcal{S}_{1/\nu^{(h)}}
\left(
\mathbf{X}
-
\boldsymbol{\Psi}
\mathbf{B}^{(h+1)}
\mathbf{A}^{(h+1)}
+
\frac{1}{\nu^{(h)}}
\mathbf{Y}^{(h)}
\right).
\label{eq:app_S_update_final}
\end{equation}


\bibliographystyle{unsrt}  
\bibliography{refs}

\end{document}